\documentclass[11pt]{article}
\usepackage{cite}
\usepackage{mdframed}
\usepackage{epsfig,subfigure}
\usepackage{amssymb}
\usepackage[fleqn]{amsmath}
\usepackage{color}
\usepackage{arydshln}
\def\QED{\mbox{\rule[0pt]{1.5ex}{1.5ex}}}

\usepackage{graphicx}
\usepackage{color}
\usepackage[dvipsnames]{xcolor}

\definecolor{armygreen}{rgb}{0.29, 0.33, 0.13}

\usepackage{amssymb}
\usepackage{amsmath,amsfonts,amssymb}
\usepackage{verbatim}
\usepackage{stfloats}
\usepackage[bookmarks=false]{}
\usepackage{algorithm}
\usepackage{algcompatible}
\usepackage{tikz,pgfplots,tikzpeople}
\usepackage{subfig}

\newtheorem{theorem}{Theorem}

\newtheorem{definition}{Definition}
\newtheorem{lemma}{Lemma}

\newcommand\blfootnote[1]{%
  \begingroup
  \renewcommand\thefootnote{}\footnote{#1}%
  \addtocounter{footnote}{-1}%
  \endgroup
}

\begin{document}
\date{}

\title{
On Extremal Rates of Storage over Graphs
}
\author{\normalsize Zhou Li and Hua Sun \\
}

\maketitle

\blfootnote{
Zhou Li (email: zhouli@my.unt.edu) and Hua Sun (email: hua.sun@unt.edu) are with the Department of Electrical Engineering at the University of North Texas. }

\maketitle

\begin{abstract}
A storage code over a graph maps $K$ independent source symbols, each of $L_w$ bits, to $N$ coded symbols, each of $L_v$ bits, such that each coded symbol is stored in a node of the graph and each edge of the graph is associated with one source symbol. From a pair of nodes connected by an edge, the source symbol that is associated with the edge can be decoded. 
The ratio $L_w/L_v$ is called the symbol rate of a storage code and the highest symbol rate is called the capacity. We show that the three highest capacity values of storage codes over graphs are $2, 3/2, 4/3$. We characterize all graphs over which the storage code capacity is $2$ and $3/2$, and for capacity value of $4/3$, necessary condition and sufficient condition (that do not match) on the graphs are given. 
\end{abstract}

\newpage

\allowdisplaybreaks
\section{Introduction}

Motivated by the heterogeneity of modern distributed storage systems, a storage code problem over graphs is introduced in \cite{Li_Sun_SecureStorage, Sahraei_Gastpar}, where a storage code maps $K$ independent source symbols, $W_1, \cdots, W_K$ to $N$ coded symbols, $V_1, \cdots, V_N$, and the coded symbols are stored in the node set of a graph $\{V_1, \cdots, V_N\}$ (so that $V_n$ denotes both the coded symbol and the node). The heterogeneous data recovery pattern is captured by the edges of the graph, where each edge $\{V_i, V_j\}$ is associated with one source symbol $W_k$ and from $(V_i, V_j)$, we can decode $W_k$. As the structure of the graph can be very diverse, versatile distributed storage and data access requirements can be accommodated. An example of the storage code problem over a graph is given in Fig.~\ref{fig:prob}. The metric of pursuit is the capacity $C$ of a storage code over a graph, i.e., the highest possible symbol rate, defined as $L_w/L_v$, where $L_w (L_v)$ is the number of bits contained in each source (coded) symbol and $L_w/L_v$ represents the number of source symbol bits reliably stored in each coded symbol bit.

\vspace{0.1in}
\tikzset{
    photon/.style={decorate, decoration={snake,
      amplitude = 0.5mm,
      segment length = 3mm}, draw=red},
    electron/.style={draw=blue, postaction={decorate},
    decoration={markings,mark=at position .55 with {\arrow[draw=blue]{>}}}},
    gluon/.style={decorate, draw=blue,
        decoration={coil,amplitude=1.5pt, segment length=4pt}} 
}
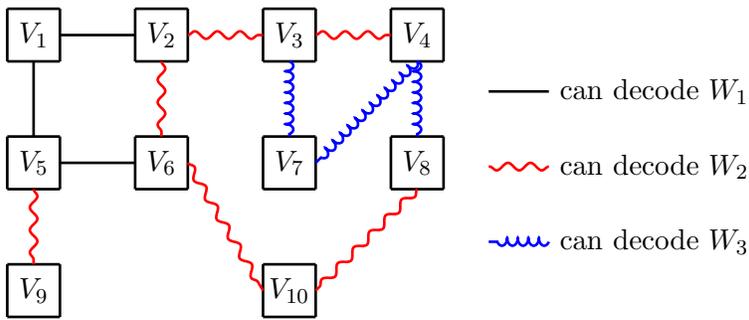
\begin{figure}[H]
\begin{center}
\begin{tikzpicture}

\filldraw (1.35,2.25) node {$V_1$};
\filldraw (3.05,2.25) node {$V_2$};
\filldraw (4.75,2.25) node {$V_3$};
\filldraw (6.45,2.25) node {$V_4$};

\filldraw (1.35,0.55) node {$V_5$};
\filldraw (3.05,0.55) node {$V_6$};
\filldraw (4.75,0.55) node {$V_7$};
\filldraw (6.45,0.55) node {$V_8$};

\filldraw (1.35,-1.15) node {$V_9$};
\filldraw (4.75,-1.15) node {$V_{10}$};

\draw[black, line width=1] (1,-1.5) -- (1,-0.8);
\draw[black, line width=1] (1,-1.5) -- (1.7,-1.5);
\draw[black, line width=1] (1,-0.8) -- (1.7,-0.8);
\draw[black, line width=1] (1.7,-1.5) -- (1.7,-0.8);

\draw[black, line width=1] (1,0.2) -- (1,0.9);
\draw[black, line width=1] (1,0.2) -- (1.7,0.2);
\draw[black, line width=1] (1,0.9) -- (1.7,0.9);
\draw[black, line width=1] (1.7,0.2) -- (1.7,0.9);

\draw[black, line width=1] (2.7,0.2) -- (2.7,0.9);
\draw[black, line width=1] (2.7,0.2) -- (3.4,0.2);
\draw[black, line width=1] (2.7,0.9) -- (3.4,0.9);
\draw[black, line width=1] (3.4,0.2) -- (3.4,0.9);

\draw[black, line width=1] (1,1.9) -- (1,2.6);
\draw[black, line width=1] (1,1.9) -- (1.7,1.9);
\draw[black, line width=1] (1,2.6) -- (1.7,2.6);
\draw[black, line width=1] (1.7,1.9) -- (1.7,2.6);

\draw[black, line width=1] (2.7,1.9) -- (2.7,2.6);
\draw[black, line width=1] (2.7,1.9) -- (3.4,1.9);
\draw[black, line width=1] (2.7,2.6) -- (3.4,2.6);
\draw[black, line width=1] (3.4,1.9) -- (3.4,2.6);

\draw[black, line width=1] (4.4,-1.5) -- (4.4,-0.8);
\draw[black, line width=1] (4.4,-1.5) -- (5.1,-1.5);
\draw[black, line width=1] (4.4,-0.8) -- (5.1,-0.8);
\draw[black, line width=1] (5.1,-1.5) -- (5.1,-0.8);

\draw[black, line width=1] (4.4,0.2) -- (4.4,0.9);
\draw[black, line width=1] (4.4,0.2) -- (5.1,0.2);
\draw[black, line width=1] (4.4,0.9) -- (5.1,0.9);
\draw[black, line width=1] (5.1,0.2) -- (5.1,0.9);

\draw[black, line width=1] (6.1,0.2) -- (6.1,0.9);
\draw[black, line width=1] (6.1,0.2) -- (6.8,0.2);
\draw[black, line width=1] (6.1,0.9) -- (6.8,0.9);
\draw[black, line width=1] (6.8,0.2) -- (6.8,0.9);

\draw[black, line width=1] (4.4,1.9) -- (4.4,2.6);
\draw[black, line width=1] (4.4,1.9) -- (5.1,1.9);
\draw[black, line width=1] (4.4,2.6) -- (5.1,2.6);
\draw[black, line width=1] (5.1,1.9) -- (5.1,2.6);

\draw[black, line width=1] (6.1,1.9) -- (6.1,2.6);
\draw[black, line width=1] (6.1,1.9) -- (6.8,1.9);
\draw[black, line width=1] (6.1,2.6) -- (6.8,2.6);
\draw[black, line width=1] (6.8,1.9) -- (6.8,2.6);

\draw[black, line width=1] (1.7,2.25) -- (2.7,2.25);
\draw[black, line width=1] (1.35,1.9) -- (1.35,0.9);
\draw[black, line width=1] (1.7,0.55) -- (2.7,0.55);

\draw[red, photon, line width=1] (1.35,0.2) -- (1.35,-0.8);
\draw[red, photon, line width=1] (3.05,1.9) -- (3.05,0.9);

\draw[red, photon, line width=1] (3.4,2.25) -- (4.4,2.25);
\draw[red, photon, line width=1] (5.1,2.25) -- (6.1,2.25);

\draw[red, photon, line width=1] (3.4,0.55) -- (4.4,-1.15);
\draw[red, photon, line width=1] (6.45,0.2) -- (5.1,-1.15);

\draw[blue, gluon, line width=1] (4.75,1.9) -- (4.75,0.9);
\draw[blue, gluon, line width=1] (6.45,1.9) -- (6.45,0.9);

\draw[blue, gluon, line width=1] (6.45,1.9) -- (5.1,0.55);

\draw[black, line width=1] (8.2,1.5) -- (7.4,1.5);
\draw[red, photon, line width=1] (8.2,0.5) -- (7.4,0.5);
\draw[blue, gluon, line width=1] (8.2,-0.5) -- (7.4,-0.5);

\filldraw (9.6,1.5) node {can decode $W_1$};
\filldraw (9.6,0.5) node {can decode $W_2$};
\filldraw (9.6,-0.5) node {can decode $W_3$};

\end{tikzpicture}

\end{center}
\vspace{-0.1in}
  \caption{\small An example graph of a storage code problem with $K=3$ source symbols and $N=10$ coded symbols, whose capacity turns out to be $4/3$ (refer to Theorem \ref{thm:sufficient2}. See Fig.~\ref{fig:sufficient2} for a code construction).
  }
  \label{fig:prob}
\end{figure}

The graph based storage code problem is not new in the sense that it can be equivalently transformed to a network coding problem \cite{Li_Sun_SecureStorage, Sahraei_Gastpar, Y_ITNC, combination_network} and adding further security constraints (i.e., beyond desired data decodability, leakage about other source symbols is prevented), it is intimately related to conditional disclosure of secrets \cite{SymPIR, Li_Sun_CDS, Li_Sun_linearCDS, Wang_Ulukus_CDMS} and secret sharing \cite{Brickell_Davenport_Ideal, Sun_Shieh_Graph}. What is new is the view brought by \cite{Li_Sun_SecureStorage} - finding extremal networks/graphs. Instead of first fixing the network/graph and then finding its highest rate, we focus on the extremal (highest) 
capacity values and aim to find the networks/graphs whose capacity is equal to the extremal values (see Fig.~\ref{fig:approach}). This complementary view is useful in identifying critical combinatorial graph structures that limit the rate and in separating more tractable graph classes in terms of capacity characterization. Considering that networks are becoming more and more heterogeneous and solving each network instance becomes infeasible and impossible (as hard instances that require non-linear codes for achievability or non-Shannon information inequalities for converse are well known \cite{DFZ_Matroids, Kamath_Anantharam_Tse_Wang, Sun_Jafar_nonshannon}), this extremal rate and extremal network approach might be a fruitful direction to produce new results and insights.

\vspace{0.1in}
\tikzset{
    photon/.style={decorate, decoration={snake,
      amplitude = 0.5mm,
      segment length = 3mm}, draw=red},
    electron/.style={draw=blue, postaction={decorate},
    decoration={markings,mark=at position .55 with {\arrow[draw=blue]{>}}}},
    gluon/.style={decorate, draw=blue,
        decoration={coil,amplitude=1.5pt, segment length=4pt}} 
}
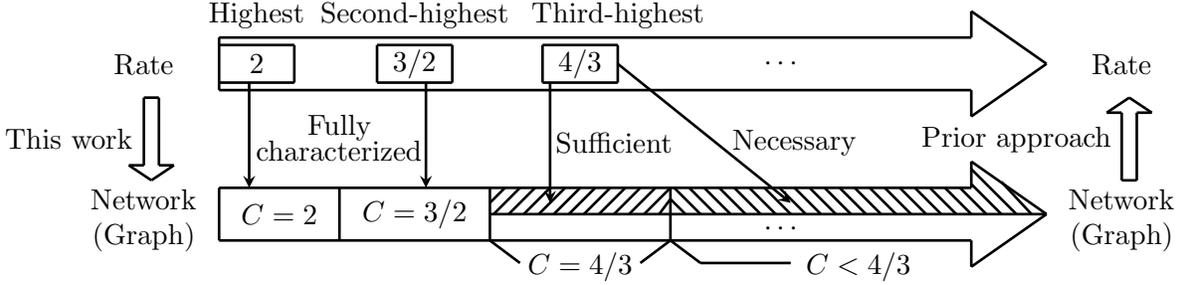
\begin{figure}[H]
\begin{center}
\begin{tikzpicture}

\filldraw (0,0.55) node {Network};
\filldraw (0,0.05) node {(Graph)};
\filldraw (0,2.35) node {Rate};
\filldraw (13,0.55) node {Network};
\filldraw (13,0.05) node {(Graph)};
\filldraw (13,2.35) node {Rate};
\filldraw (-1,1.35) node {This work};
\filldraw (11.6,1.35) node {Prior approach};

\filldraw (1.5,3) node {Highest};
\filldraw (1.5,2.35) node {$2$};

\filldraw (3.6,3) node {Second-highest};
\filldraw (3.6,2.35) node {$3/2$};

\filldraw (6.3,3) node {Third-highest};
\filldraw (5.8,2.35) node {$4/3$};

\filldraw (8.5,2.35) node {$\cdots$};

\draw[black, line width=1] (13,0.8) -- (13.2,0.8);
\draw[black, line width=1] (13,0.8) -- (13,1.7);
\draw[black, line width=1] (13.2,0.8) -- (13.2,1.7);
\draw[black, line width=1] (13.1,1.9) -- (12.8,1.7);
\draw[black, line width=1] (13.1,1.9) -- (13.4,1.7);
\draw[black, line width=1] (12.8,1.7) -- (13,1.7);
\draw[black, line width=1] (13.4,1.7) -- (13.2,1.7);

\draw[black, line width=1] (1,2) -- (11,2);
\draw[black, line width=1] (1,2.7) -- (11,2.7);

\draw[black, line width=1] (11,2.7) -- (11,3.05);
\draw[black, line width=1] (11,2) -- (11,1.65);
\draw[black, line width=1] (12,2.35) -- (11,3.05);
\draw[black, line width=1] (12,2.35) -- (11,1.65);

\draw[black, line width=1] (1,2) -- (1,2.7);
\draw[black, line width=1] (2,2.1) -- (2,2.6);
\draw[black, line width=1] (1,2.1) -- (2,2.1);
\draw[black, line width=1] (1,2.6) -- (2,2.6);

\draw[black, line width=1] (3.1,2.1) -- (3.1,2.6);
\draw[black, line width=1] (4.1,2.1) -- (4.1,2.6);
\draw[black, line width=1] (3.1,2.1) -- (4.1,2.1);
\draw[black, line width=1] (3.1,2.6) -- (4.1,2.6);

\draw[black, line width=1] (5.3,2.1) -- (5.3,2.6);
\draw[black, line width=1] (6.3,2.1) -- (6.3,2.6);
\draw[black, line width=1] (5.3,2.1) -- (6.3,2.1);
\draw[black, line width=1] (5.3,2.6) -- (6.3,2.6);

\draw[black, line width=1] (0,1.9) -- (0.2,1.9);
\draw[black, line width=1] (0,1.9) -- (0,1);
\draw[black, line width=1] (0.2,1.9) -- (0.2,1);
\draw[black, line width=1] (0.1,0.8) -- (0.4,1);
\draw[black, line width=1] (0.1,0.8) -- (-0.2,1);
\draw[black, line width=1] (-0.2,1) -- (0,1);
\draw[black, line width=1] (0.2,1) -- (0.4,1);

\draw[black, line width=1] (1,0) -- (11,0);
\draw[black, line width=1] (1,0.7) -- (11,0.7);

\draw[black, line width=1] (1,0) -- (1,0.7);
\draw[black, line width=1] (2.6,0) -- (2.6,0.7);
\draw[black, line width=1] (4.6,0) -- (4.6,0.7);
\draw[black, line width=1] (7,0) -- (7,0.7);
\draw[black, line width=1] (4.6,0.35) -- (12,0.35);

\draw[black, line width=1] (4.6,0.5) -- (4.8,0.7);
\draw[black, line width=1] (4.65,0.35) -- (5,0.7);
\draw[black, line width=1] (4.85,0.35) -- (5.2,0.7);
\draw[black, line width=1] (5.05,0.35) -- (5.4,0.7);
\draw[black, line width=1] (5.25,0.35) -- (5.6,0.7);
\draw[black, line width=1] (5.45,0.35) -- (5.8,0.7);
\draw[black, line width=1] (5.65,0.35) -- (6,0.7);
\draw[black, line width=1] (5.85,0.35) -- (6.2,0.7);
\draw[black, line width=1] (6.05,0.35) -- (6.4,0.7);
\draw[black, line width=1] (6.25,0.35) -- (6.6,0.7);
\draw[black, line width=1] (6.45,0.35) -- (6.8,0.7);
\draw[black, line width=1] (6.65,0.35) -- (7,0.7);
\draw[black, line width=1] (6.85,0.35) -- (7,0.55);

\draw[black, line width=1] (7,0.55) -- (7.2,0.35);
\draw[black, line width=1] (7.05,0.7) -- (7.4,0.35);
\draw[black, line width=1] (7.25,0.7) -- (7.6,0.35);
\draw[black, line width=1] (7.45,0.7) -- (7.8,0.35);
\draw[black, line width=1] (7.65,0.7) -- (8,0.35);
\draw[black, line width=1] (7.85,0.7) -- (8.2,0.35);
\draw[black, line width=1] (8.05,0.7) -- (8.4,0.35);
\draw[black, line width=1] (8.25,0.7) -- (8.6,0.35);
\draw[black, line width=1] (8.45,0.7) -- (8.8,0.35);
\draw[black, line width=1] (8.65,0.7) -- (9,0.35);
\draw[black, line width=1] (8.85,0.7) -- (9.2,0.35);
\draw[black, line width=1] (9.05,0.7) -- (9.4,0.35);
\draw[black, line width=1] (9.25,0.7) -- (9.6,0.35);
\draw[black, line width=1] (9.45,0.7) -- (9.8,0.35);
\draw[black, line width=1] (9.65,0.7) -- (10,0.35);
\draw[black, line width=1] (9.85,0.7) -- (10.2,0.35);
\draw[black, line width=1] (10.05,0.7) -- (10.4,0.35);
\draw[black, line width=1] (10.25,0.7) -- (10.6,0.35);
\draw[black, line width=1] (10.45,0.7) -- (10.8,0.35);
\draw[black, line width=1] (10.65,0.7) -- (11,0.35);
\draw[black, line width=1] (10.85,0.7) -- (11.2,0.35);
\draw[black, line width=1] (11.0,0.75) -- (11.4,0.35);
\draw[black, line width=1] (11.0,0.95) -- (11.6,0.35);
\draw[black, line width=1] (11.4,0.78) -- (11.8,0.35);

\draw[black, line width=1] (4.6,0) -- (5,-0.3);
\draw[black, line width=1] (7,0) -- (6.6,-0.3);
\draw[black, line width=1] (7,0) -- (7.4,-0.3);
\draw[black, line width=1] (8.6,-0.3) -- (7.4,-0.3);

\draw[black, line width=1] (11,0.7) -- (11,1.05);
\draw[black, line width=1] (11,0) -- (11,-0.35);
\draw[black, line width=1] (12,0.35) -- (11,1.05);
\draw[black, line width=1] (12,0.35) -- (11,-0.35);

\filldraw (1.8,0.35) node {$C = 2$};
\filldraw (3.6,0.35) node {$C = 3/2$};
\filldraw (5.8,-0.35) node {$C = 4/3$};
\filldraw (9.5,-0.35) node {$C < 4/3$};
\filldraw (8.5,0.15) node {$\cdots$};
;

\draw[-stealth][black, line width=1] (1.4,2.1) -- (1.4,0.7);
\filldraw (2.6,1.5) node {Fully};
\filldraw (2.6,1.2) node {characterized};

\draw[-stealth][black, line width=1] (3.75,2.1) -- (3.75,0.7);

\draw[-stealth][black, line width=1] (5.4,2.1) -- (5.4,0.5);
\filldraw (6.25,1.3) node {Sufficient};

\draw[-stealth][black, line width=1] (6.3,2.35) -- (8.6,0.5);
\filldraw (8.65,1.3) node {Necessary};

\end{tikzpicture}

\end{center}
\vspace{-0.1in}
  \caption{\small The extremal rate and network approach of this work and results obtained.
  }
  \label{fig:approach}
\end{figure}

In this work, we start from the highest possible capacity values and 
for the two highest rates - $2$ and $3/2$, all extremal graphs with corresponding extremal capacity values are easily characterized. For extremal rate of $2$, absolute no interference is allowed as $L_w = 2L_v$, i.e., a pair of nodes can just store the desired source symbols. As long as there exists interference, the maximal capacity value drops to $3/2$, the next extremal rate, and all storage code instances with capacity $3/2$ only require intra-source symbol coding, i.e., mixing of symbols from the same source symbol. When rate of $3/2$ cannot be achieved, the next highest capacity value is shown to be $4/3$, which is our main focus and the corresponding graphs turn out to be highly technical. We identify necessary condition (converse required) and sufficient condition (achievability provided) for graphs with storage code capacity $4/3$ (see Fig.~\ref{fig:approach}). The converse is based on delicate arguments on the intimate relation between the maximum amount of interference (undesired source symbols) allowed and the minimum amount of desired source symbols needed. The achievable scheme uses vector linear codes that carefully control the alignment of interfering source symbols and the independence of desired source symbols. The conditions are stated in terms of the presence or absence of critical nodes and edges of the graph, whose combinatorial structure places constraints on the code rate.

\section{Problem Statement and Definitions}\label{sec:model}
Consider $K$ independent uniform source symbols $W_1, \cdots, W_K$ of size $L_w$ bits each.
\begin{eqnarray}
&& H(W_1, \cdots, W_K) = H(W_1) + \cdots + H(W_K), \notag\\
&& L_w = H(W_1) = \cdots = H(W_K). \label{h1}
\end{eqnarray}

Consider $N$ coded symbols $V_1, \cdots, V_N$, each of $L_v$ bits. Our interest lies in the relative size of $L_w, L_v$ (see (\ref{rate})) and coding over arbitrary finite fields is allowed, so $L_w, L_v$ can take arbitrarily large values (that are not necessarily integers).

The source symbol recoverability constraint on the coded symbols is specified by a graph $G = (\mathcal{V}, \mathcal{E}, t)$, where the node\footnote{Note that we abuse the notation by using $V_n$ to denote both a coded symbol and a node of the graph, which will not cause confusion.} set $\mathcal{V} = \{V_1, \cdots, V_N\}$, the edge set $\mathcal{E}$ is a set of unordered pairs from $\mathcal{V}$, and the function $t$ associates each edge $\{V_i, V_j\} \in \mathcal{E}$ with a source symbol $W_k, k \in \{1,2,\cdots,K\}\triangleq [K]$, i.e., $t(\{V_i,V_j\}) = W_k$. For each edge $\{V_i, V_j\} \in \mathcal{E}$ such that $t(\{V_i, V_j\}) = W_k$, we can decode $W_k$ with no error, i.e., 
\begin{eqnarray}
 H(W_k | V_i, V_j ) = 0 ~\mbox{if}~ t(\{V_i, V_j\}) = W_k. \label{dec} 
\end{eqnarray}
Isolated nodes are trivial as they are not connected to any edges and thus involve no constraints. Without loss of generality, we assume in this work that any graph contains no isolated nodes.

A mapping from the source symbols $W_1,\cdots,W_K$ to the coded symbols $V_1, \cdots, V_N$ that satisfies the decoding constraint (\ref{dec}) specified by a graph $G = (\mathcal{V}, \mathcal{E}, t)$ is called a storage code. The (achievable) symbol rate of a storage code is defined as
\begin{eqnarray}
R \triangleq \frac{L_w}{L_v} \label{rate}
\end{eqnarray}
and the supremum of symbol rate is called the capacity, $C \triangleq 
\sup_{L_w} L_w/L_v = \lim_{L_w\rightarrow \infty} L_w/L_v$, as block codes are allowed.

Next we introduce some graph definitions to facilitate the presentation of our results.

\subsection{Graph Definitions}

\begin{definition}[$W_k$-Edge, $W_k$-Path, and $W_k$-Component]
An edge that is associated with $W_k$ is called a $W_k$-edge. A sequence of distinct connecting $W_k$-edges is called a $W_k$-path. 
A $W_k$-component is a maximal subgraph wherein every edge is a $W_k$-edge and every two nodes are connected by a $W_k$-path (an isolated node is defined as a trivial component). 
\end{definition}

For example, in Fig.~\ref{fig:prob}, $\{V_1, V_2\}$ (also all solid black edges) is a $W_1$-edge; the sequence of $W_1$-edges $(\{V_2, V_1\}, \{V_1, V_5\}, \{V_5, V_6\})$ is a $W_1$-path and also a $W_1$-component. 

\begin{definition}[Internal Edge and Residing Path]
A $W_k$-edge that connects two nodes (say $V_i, V_j$) in a $W_{k'}$-path, $k' \neq k$ is said to be internal and the $W_{k'}$-path with end nodes $V_i, V_j$ is called the residing path of the internal $W_k$-edge $\{V_i, V_j\}$.
\end{definition}

For example, in Fig.~\ref{fig:prob}, the $W_2$-edge $\{V_2, V_6\}$ is an internal edge as it connects two nodes $V_2, V_6$ in the $W_1$-path $(\{V_2, V_1\}, \{V_1, V_5\}, \{V_5, V_6\})$, which is then its residing path.

\begin{definition}[$M$-Color Node] \label{def:cha}
A node whose connected edges are associated with $M$ different source symbols is called an $M$-color node. 
\end{definition}

For example, in Fig.~\ref{fig:prob}, $V_1, V_9$ are $1$-color nodes and $V_5, V_6$ are $2$-color nodes.

We need to further distinguish two types of $2$-color nodes, defined as follows.

\begin{definition}[Normal $2$-Color Node and $W_k$-Special $2$-Color Node] 
For a $2$-color node $V$ that is connected to $W_k$-edges and $W_{k'}$-edges, $k\neq k'$, if the nodes connected to $V$ through $W_k$-edges are all $1$-color, then $V$ is called a $W_k$-special $2$-color node (or just a special $2$-color node when $W_k$ does not need to be highlighted). A $2$-color node that is not special is said to be normal. 
\end{definition}

For example, in Fig.~\ref{fig:prob}, the $2$-color node $V_5$ is $W_2$-special as $V_9$ is the only node that is connected to $V_5$ through $W_2$-edges and $V_9$ is $1$-color; the $2$-color node $V_6$ is normal as it is connected to a $2$-color node $V_2$ through a $W_2$-edge and is connected to a $2$-color node $V_5$ through a $W_1$-edge.

\begin{definition}[Graph Class $\mathcal{G}_{C=R^*}, \mathcal{G}_{C\geq R^*}, \mathcal{G}_{C < R^*}$]
The set of graphs whose storage code capacity is equal to$\backslash$no smaller than$\backslash$strictly smaller than $R^*$ is denoted by $\mathcal{G}_{C=R^*} \backslash \mathcal{G}_{C\geq R^*} \backslash \mathcal{G}_{C < R^*}$.
\end{definition}

\section{Results}

Our results are presented in this section, along with illustrative examples and observations.

\subsection{Extremal Graphs with Storage Code Capacity $2, 3/2$: $\mathcal{G}_{C = 2}, \mathcal{G}_{C = 3/2}$}

The three highest extremal capacity values and the full extremal graph characterization for the two highest extremal capacity values are established in the following theorem.

\begin{theorem}\label{thm:C2} [$\mathcal{G}_{C = 2}, \mathcal{G}_{C = 3/2}$]
The three highest storage code capacity values are $2, 3/2, 4/3$. The storage code capacity of a graph is equal to $2$ ($G \in \mathcal{G}_{C=2}$) if and only if every node is $1$-color. 
The storage code capacity of a graph is equal to $3/2$ ($G \in \mathcal{G}_{C=3/2}$) if and only if all nodes are $2$-color or $1$-color (and $2$-color nodes exist) and there are no connected $2$-color nodes.
\end{theorem}

The proof of Theorem \ref{thm:C2} is fairly straightforward and is deferred to Section \ref{sec:C2}. An example of the achievable scheme (code construction) is shown in Fig.~\ref{fig:C2}.(a) and Fig.~\ref{fig:C2}.(b). An example graph that does not belong to $\mathcal{G}_{C=2} \cup \mathcal{G}_{C=3/2}$ is shown in Fig.~\ref{fig:C2}.(c). An intuitive explanation on why the rate is upper bounded by $4/3$ is as follows. $V_3$ can at most contribute $L_v$ bits of information about $W_2$. $\{V_1, V_3\}$ is a $W_2$-edge so that $V_1$ has to provide at least the remaining $L_w - L_v$ bits of information about $W_2$, leaving at most $L_v - (L_w - L_v) = 2L_v - L_w$ bits of room for $W_1$. The same reasoning applies to $V_2$. Finally, $\{V_1, V_2\}$ is a $W_1$-edge so that the size of the remaining room must accommodate the $L_w$ bits of $W_1$, i.e., $2(2L_v - L_w) \geq L_w$ so that $R = L_w/L_v \leq 4/3$.

\tikzset{
    photon/.style={decorate, decoration={snake,
      amplitude = 0.5mm,
      segment length = 3mm}, draw=red},
    electron/.style={draw=blue, postaction={decorate},
        decoration={markings,mark=at position .55 with {\arrow[draw=blue]{>}}}},
    gluon/.style={decorate, draw=blue,
        decoration={coil,amplitude=1.5pt, segment length=4pt}} 
}
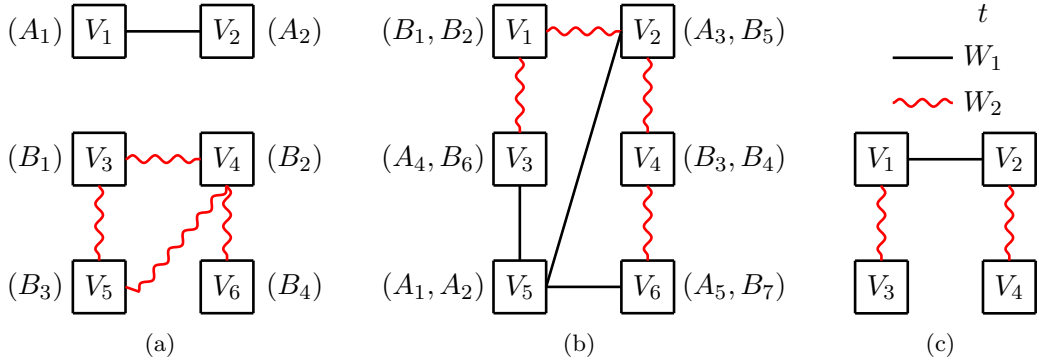
\begin{figure}[H]
\begin{center}
\subfigure[]{
\begin{tikzpicture}
\filldraw (1.35,2.25) node {$V_1$};
\filldraw (3.05,2.25) node {$V_2$};

\filldraw (1.35,0.55) node {$V_3$};
\filldraw (3.05,0.55) node {$V_4$};

\filldraw (1.35,-1.15) node {$V_5$};
\filldraw (3.05,-1.15) node {$V_6$};

\draw[black, line width=1] (1,-1.5) -- (1,-0.8);
\draw[black, line width=1] (1,-1.5) -- (1.7,-1.5);
\draw[black, line width=1] (1,-0.8) -- (1.7,-0.8);
\draw[black, line width=1] (1.7,-1.5) -- (1.7,-0.8);

\draw[black, line width=1] (2.7,-1.5) -- (2.7,-0.8);
\draw[black, line width=1] (2.7,-1.5) -- (3.4,-1.5);
\draw[black, line width=1] (2.7,-0.8) -- (3.4,-0.8);
\draw[black, line width=1] (3.4,-1.5) -- (3.4,-0.8);

\draw[black, line width=1] (1,0.2) -- (1,0.9);
\draw[black, line width=1] (1,0.2) -- (1.7,0.2);
\draw[black, line width=1] (1,0.9) -- (1.7,0.9);
\draw[black, line width=1] (1.7,0.2) -- (1.7,0.9);

\draw[black, line width=1] (2.7,0.2) -- (2.7,0.9);
\draw[black, line width=1] (2.7,0.2) -- (3.4,0.2);
\draw[black, line width=1] (2.7,0.9) -- (3.4,0.9);
\draw[black, line width=1] (3.4,0.2) -- (3.4,0.9);

\draw[black, line width=1] (1,1.9) -- (1,2.6);
\draw[black, line width=1] (1,1.9) -- (1.7,1.9);
\draw[black, line width=1] (1,2.6) -- (1.7,2.6);
\draw[black, line width=1] (1.7,1.9) -- (1.7,2.6);

\draw[black, line width=1] (2.7,1.9) -- (2.7,2.6);
\draw[black, line width=1] (2.7,1.9) -- (3.4,1.9);
\draw[black, line width=1] (2.7,2.6) -- (3.4,2.6);
\draw[black, line width=1] (3.4,1.9) -- (3.4,2.6);

\draw[black, line width=1] (1.7,2.25) -- (2.7,2.25);

\draw[red, photon, line width=1] (1.7,0.55) -- (2.7,0.55);
\draw[red, photon, line width=1] (3.05,0.2) -- (1.7,-1.15);
\draw[red, photon, line width=1] (1.35,0.2) -- (1.35,-0.8);
\draw[red, photon, line width=1] (3.05,0.2) -- (3.05,-0.8);

\filldraw (0.5,2.25) node {$(A_1)$};
\filldraw (3.95,2.25) node {$(A_2)$};

\filldraw (0.5,0.55) node {$(B_1)$};
\filldraw (3.95,0.55) node {$(B_2)$};

\filldraw (0.5,-1.15) node {$(B_3)$};
\filldraw (3.95,-1.15) node {$(B_4)$};

\end{tikzpicture}
}
\hspace{0in}
\subfigure[]{ 
\begin{tikzpicture}
\filldraw (1.35,2.25) node {$V_1$};
\filldraw (3.05,2.25) node {$V_2$};

\filldraw (1.35,0.55) node {$V_3$};
\filldraw (3.05,0.55) node {$V_4$};

\filldraw (1.35,-1.15) node {$V_5$};
\filldraw (3.05,-1.15) node {$V_6$};

\draw[black, line width=1] (1,-1.5) -- (1,-0.8);
\draw[black, line width=1] (1,-1.5) -- (1.7,-1.5);
\draw[black, line width=1] (1,-0.8) -- (1.7,-0.8);
\draw[black, line width=1] (1.7,-1.5) -- (1.7,-0.8);

\draw[black, line width=1] (2.7,-1.5) -- (2.7,-0.8);
\draw[black, line width=1] (2.7,-1.5) -- (3.4,-1.5);
\draw[black, line width=1] (2.7,-0.8) -- (3.4,-0.8);
\draw[black, line width=1] (3.4,-1.5) -- (3.4,-0.8);

\draw[black, line width=1] (1,0.2) -- (1,0.9);
\draw[black, line width=1] (1,0.2) -- (1.7,0.2);
\draw[black, line width=1] (1,0.9) -- (1.7,0.9);
\draw[black, line width=1] (1.7,0.2) -- (1.7,0.9);

\draw[black, line width=1] (2.7,0.2) -- (2.7,0.9);
\draw[black, line width=1] (2.7,0.2) -- (3.4,0.2);
\draw[black, line width=1] (2.7,0.9) -- (3.4,0.9);
\draw[black, line width=1] (3.4,0.2) -- (3.4,0.9);

\draw[black, line width=1] (1,1.9) -- (1,2.6);
\draw[black, line width=1] (1,1.9) -- (1.7,1.9);
\draw[black, line width=1] (1,2.6) -- (1.7,2.6);
\draw[black, line width=1] (1.7,1.9) -- (1.7,2.6);

\draw[black, line width=1] (2.7,1.9) -- (2.7,2.6);
\draw[black, line width=1] (2.7,1.9) -- (3.4,1.9);
\draw[black, line width=1] (2.7,2.6) -- (3.4,2.6);
\draw[black, line width=1] (3.4,1.9) -- (3.4,2.6);

\draw[red, photon, line width=1] (1.7,2.25) -- (2.7,2.25);
\draw[red, photon, line width=1] (1.35,1.9) -- (1.35,0.9);
\draw[red, photon, line width=1] (3.05,1.9) -- (3.05,0.9);
\draw[red, photon, line width=1] (3.05,0.2) -- (3.05,-0.8);

\draw[black, line width=1]  (1.7,-1.15) -- (2.7,2.25) ;
\draw[black, line width=1] (1.35,0.2) -- (1.35,-0.8);
\draw[black, line width=1] (1.7,-1.15) -- (2.7,-1.15);

 \filldraw (0.2,2.25) node {$(B_1,B_2)$};
 \filldraw (4.2,2.25) node {$(A_3,B_5)$};

 \filldraw (0.2,0.55) node {$(A_4,B_6)$};
 \filldraw (4.2,0.55) node {$(B_3,B_4)$};

 \filldraw (0.2,-1.15) node {$(A_1,A_2)$};
 \filldraw (4.2,-1.15) node {$(A_5,B_7)$};

\end{tikzpicture}
 }
\hspace{0.1in}
\subfigure[]{ 
\begin{tikzpicture}

\filldraw (3.05,2.25) node {$V_1$};
\filldraw (4.75,2.25) node {$V_2$};

\filldraw (3.05,0.55) node {$V_3$};
\filldraw (4.75,0.55) node {$V_4$};

\draw[black, line width=1] (4.4,0.2) -- (4.4,0.9);
\draw[black, line width=1] (4.4,0.2) -- (5.1,0.2);
\draw[black, line width=1] (4.4,0.9) -- (5.1,0.9);
\draw[black, line width=1] (5.1,0.2) -- (5.1,0.9);

\draw[black, line width=1] (2.7,0.2) -- (2.7,0.9);
\draw[black, line width=1] (2.7,0.2) -- (3.4,0.2);
\draw[black, line width=1] (2.7,0.9) -- (3.4,0.9);
\draw[black, line width=1] (3.4,0.2) -- (3.4,0.9);

\draw[black, line width=1] (2.7,1.9) -- (2.7,2.6);
\draw[black, line width=1] (2.7,1.9) -- (3.4,1.9);
\draw[black, line width=1] (2.7,2.6) -- (3.4,2.6);
\draw[black, line width=1] (3.4,1.9) -- (3.4,2.6);

\draw[black, line width=1] (4.4,1.9) -- (4.4,2.6);
\draw[black, line width=1] (4.4,1.9) -- (5.1,1.9);
\draw[black, line width=1] (4.4,2.6) -- (5.1,2.6);
\draw[black, line width=1] (5.1,1.9) -- (5.1,2.6);

\draw[black, line width=1] (3.4,2.25) -- (4.4,2.25);
\draw[red, photon, line width=1] (3.05,1.9) -- (3.05,0.9);
\draw[red, photon, line width=1] (4.75,1.9) -- (4.75,0.9);

\filldraw (4.4,4.2) node {$t$};
\draw[black, line width=1] (3.2,3.6) -- (4,3.6);
\filldraw (4.4,3.6) node {$W_1$};

\draw[red, photon, line width=1] (3.2,3) -- (4,3);
\filldraw (4.4,3) node {$W_2$};

\end{tikzpicture}
}
\end{center}
\vspace{-0.2in}
  \caption{\small (a) An example graph $G \in \mathcal{G}_{C=2}$. $W_1=(a_1, a_2), W_2 = (b_1,b_2)$, and each $A_i \backslash B_j$ is a generic linear combination of $(a_1, a_2) \backslash (b_1,b_2)$. 
  (b) An example graph $G \in \mathcal{G}_{C=3/2}$. $W_1=(a_1, a_2, a_3), W_2 = (b_1,b_2, b_3)$, and each $A_i \backslash B_j$ is a generic linear combination of $(a_1, a_2, a_3) \backslash (b_1,b_2, b_3)$. 
  (c) An example graph $G \in \mathcal{G}_{C < 3/2}$ where two $2$-color nodes $V_1, V_2$ are connected.
  }
  \label{fig:C2}
\end{figure}

\subsection{Extremal Graphs with 
Capacity $4/3$: $\mathcal{G}_{C = 4/3}$ with $K=2$ Source Symbols}

Next we focus on the storage code capacity value of $4/3$, whose extremal graph characterization turns out to be highly non-trivial. In this section, we consider the cases where there are $K=2$ source symbols to illustrate the results in a simpler setting and defer the generalizations to more than $2$ source symbols to the next section.

The obtained necessary and sufficient conditions are rather involved. To make the results more clear we give a summarizing chart in Fig.~\ref{fig:sum}.

\vspace{0.1in}
\tikzset{
    photon/.style={decorate, decoration={snake,
      amplitude = 0.5mm,
      segment length = 3mm}, draw=red},
    electron/.style={draw=blue, postaction={decorate},
    decoration={markings,mark=at position .55 with {\arrow[draw=blue]{>}}}},
    gluon/.style={decorate, draw=blue,
        decoration={coil,amplitude=1.5pt, segment length=4pt}} 
}
\begin{figure}[H]
\begin{center}
\begin{tikzpicture}

\draw[black, line width=1] (0,10) -- (4.7,10);
\draw[black, line width=1] (0,9.5) -- (4.7,9.5);
\draw[black, line width=1] (0,10) -- (0,9.5);
\draw[black, line width=1] (4.7,9.5) -- (4.7,10);

\filldraw (2.35,9.75) node {Is there any internal edge?};

\draw[-stealth] [black, line width=1](1,9.5) -- (0,8.5);
\draw[-stealth] [black, line width=1] (3.7,9.5) -- (4.7,8.5);

\filldraw (0,9) node {No};
\filldraw (4.7,9) node {Yes};

\draw[black, line width=1] (-2,8.5) -- (2,8.5);
\draw[black, line width=1] (-2,8) -- (2,8);
\draw[black, line width=1] (-2,8.5) -- (-2,8);
\draw[black, line width=1] (2,8) -- (2,8.5);

\filldraw (0,8.25) node {Rate $4/3$ is achievable.};
\filldraw (0,7.7) node {(Theorem 2)};

\filldraw (5.5,8.25) node {Does each residing path};
\filldraw (5.5,7.85) node {contain a $1$-color node?};

\draw[black, line width=1] (3.2,8.5) -- (7.8,8.5);
\draw[black, line width=1] (3.2,7.6) -- (7.8,7.6);
\draw[black, line width=1] (3.2,8.5) -- (3.2,7.6);
\draw[black, line width=1] (7.8,7.6) -- (7.8,8.5);

\draw[-stealth][black, line width=1] (4.2,7.6) -- (3.2,6.6);
\draw[-stealth][black, line width=1] (6.8,7.6) -- (7.8,6.6);

\filldraw (3.15,7.1) node {Yes};
\filldraw (7.8,7.1) node {No};

\draw[black, line width=1] (0,6.6) -- (4,6.6);
\draw[black, line width=1] (0,6.1) -- (4,6.1);
\draw[black, line width=1] (0,6.6) -- (0,6.1);
\draw[black, line width=1] (4,6.1) -- (4,6.6);

\filldraw (2,6.35) node {Rate $4/3$ is achievable.};
\filldraw (2,5.8) node {(Theorem 2)};

\filldraw (8.5,6.35) node {Is there any residing path that contains};
\filldraw (8.5,5.95) node {at most one special $2$-color node?};

\draw[black, line width=1] (5,6.6) -- (12,6.6);
\draw[black, line width=1] (5,5.7) -- (12,5.7);
\draw[black, line width=1] (5,6.6) -- (5,5.7);
\draw[black, line width=1] (12,5.7) -- (12,6.6);

\draw[-stealth][black, line width=1] (6,5.7) -- (5,4.7);
\draw[-stealth][black, line width=1] (11,5.7) -- (12,4.7);

\filldraw (5,5.2) node {Yes};
\filldraw (12,5.2) node {No};

\draw[black, line width=1] (1.9,4.7) -- (7,4.7);
\draw[black, line width=1] (1.9,4.2) -- (7,4.2);
\draw[black, line width=1] (1.9,4.7) -- (1.9,4.2);
\draw[black, line width=1] (7,4.2) -- (7,4.7);

\filldraw (4.45,4.45) node {Rate $4/3$ cannot be achieved.};
\filldraw (4.5,3.9) node {(Theorem 3)};

\draw[black, line width=1] (7.5,4.7) -- (14.5,4.7);
\draw[black, line width=1] (7.5,3) -- (14.5,3);
\draw[black, line width=1] (7.5,4.7) -- (7.5,3);
\draw[black, line width=1] (14.5,3) -- (14.5,4.7);

\filldraw (11,4.45) node {1. Rate $4/3$ is achievable if there is only};
\filldraw (10.75,4.05) node {one internal edge. (Theorem 4)};
\filldraw (10.9,3.65) node {2. One example where rate $4/3$ cannot };
\filldraw (10.3,3.25) node {be achieved. (Theorem 5)};

\end{tikzpicture}

\end{center}
\vspace{-0.1in}
  \caption{\small A summary of sufficient and necessary conditions of $\mathcal{G}_{C = 4/3}$ with $K=2$.
  }
  \label{fig:sum}
\end{figure}
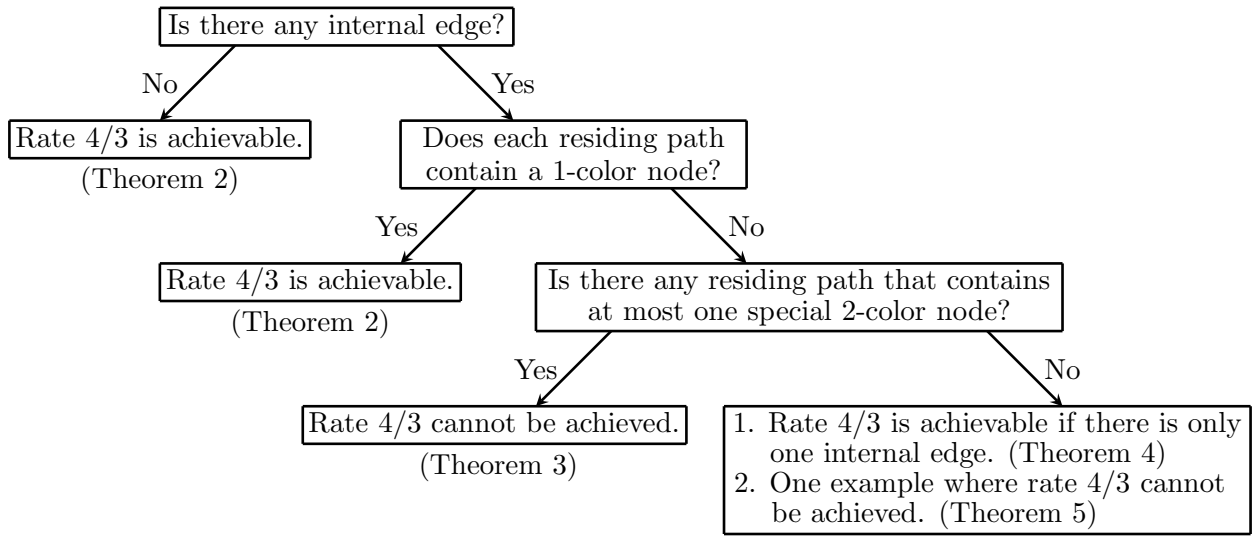

\subsubsection{Sufficient Condition: Internal Edge and $1$-Color Node}

A crucial graphic structure for the achievability of rate $4/3$ is the absence of internal edges (or when they exist, the presence of $1$-color nodes in their residing paths). This result is stated in the following theorem.

\begin{theorem}\label{thm:sufficient} [Sufficient Condition of $\mathcal{G}_{C=4/3}$]
With $K=2$ source symbols, a graph $G \in \mathcal{G}_{C \geq 4/3}$ if $G$ contains no internal edge or for any internal edge, its residing path contains a $1$-color node.
\end{theorem}

The proof of Theorem \ref{thm:sufficient} is presented in Section \ref{sec:sufficient}. To illustrate the idea, two examples are shown in Fig.~\ref{fig:sufficient}, where Example $(a)$ contains no internal edge; Example $(b)$ contains two internal edges $\{V_2, V_3\}$ and $\{V_4, V_5\}$. Internal $W_2$-edge $\{V_2, V_3\}$ resides in $W_1$-path $(\{V_2, V_1\}, \{V_1, V_3\})$, which contains $1$-color node $V_1$ and internal $W_1$-edge $\{V_3, V_5\}$ resides in $W_2$-path $(\{V_3, V_2\}, \{V_2, V_4\},$ $\{V_4, V_5\})$, which contains $1$-color node $V_4$. So the condition of Theorem \ref{thm:sufficient} is satisfied and rate $4/3$ is achievable. We next explain how to construct the code.

\tikzset{
    photon/.style={decorate, decoration={snake,
      amplitude = 0.5mm,
      segment length = 3mm}, draw=red},
    electron/.style={draw=blue, postaction={decorate},
        decoration={markings,mark=at position .55 with {\arrow[draw=blue]{>}}}},
    gluon/.style={decorate, draw=blue,
        decoration={coil,amplitude=1.5pt, segment length=4pt}} 
}
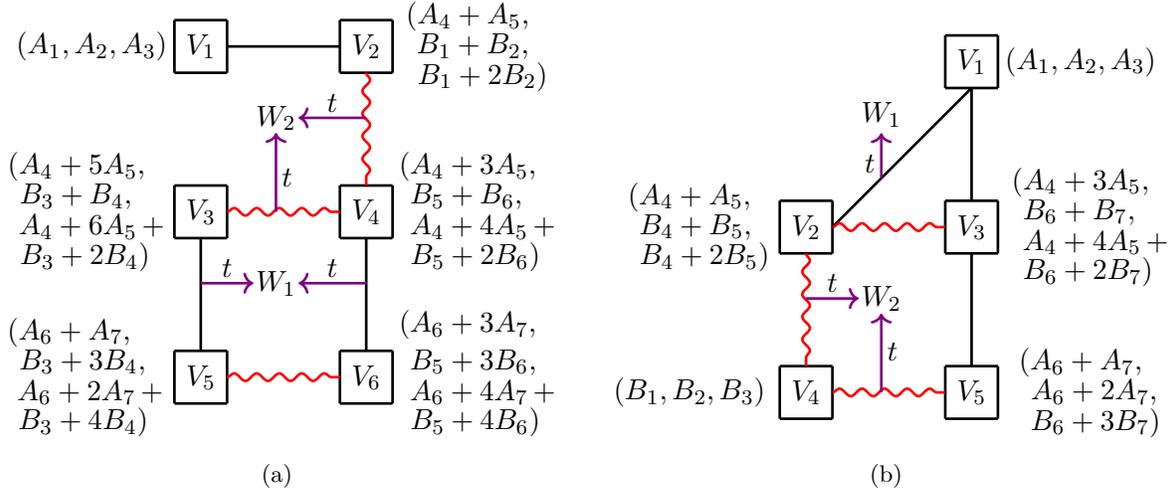
\begin{figure}[H]
\begin{center}
\subfigure[]{
\begin{tikzpicture}

\filldraw (1.35,3.25) node {$V_1$};
\filldraw (3.55,3.25) node {$V_2$};
\filldraw (1.35,1.05) node {$V_3$};
\filldraw (3.55,1.05) node {$V_4$};
\filldraw (1.35,-1.15) node {$V_5$};
\filldraw (3.55,-1.15) node {$V_6$};

\draw[black, line width=1] (1,-1.5) -- (1,-0.8);
\draw[black, line width=1] (1,-1.5) -- (1.7,-1.5);
\draw[black, line width=1] (1,-0.8) -- (1.7,-0.8);
\draw[black, line width=1] (1.7,-1.5) -- (1.7,-0.8);

\draw[black, line width=1] (3.2,-1.5) -- (3.2,-0.8);
\draw[black, line width=1] (3.2,-1.5) -- (3.9,-1.5);
\draw[black, line width=1] (3.2,-0.8) -- (3.9,-0.8);
\draw[black, line width=1] (3.9,-1.5) -- (3.9,-0.8);

\draw[black, line width=1] (1,0.7) -- (1,1.4);
\draw[black, line width=1] (1,0.7) -- (1.7,0.7);
\draw[black, line width=1] (1,1.4) -- (1.7,1.4);
\draw[black, line width=1] (1.7,0.7) -- (1.7,1.4);

\draw[black, line width=1] (3.2,0.7) -- (3.2,1.4);
\draw[black, line width=1] (3.2,0.7) -- (3.9,0.7);
\draw[black, line width=1] (3.2,1.4) -- (3.9,1.4);
\draw[black, line width=1] (3.9,0.7) -- (3.9,1.4);

\draw[black, line width=1] (1,2.9) -- (1,3.6);
\draw[black, line width=1] (1,2.9) -- (1.7,2.9);
\draw[black, line width=1] (1,3.6) -- (1.7,3.6);
\draw[black, line width=1] (1.7,2.9) -- (1.7,3.6);

\draw[black, line width=1] (3.2,2.9) -- (3.2,3.6);
\draw[black, line width=1] (3.2,2.9) -- (3.9,2.9);
\draw[black, line width=1] (3.2,3.6) -- (3.9,3.6);
\draw[black, line width=1] (3.9,2.9) -- (3.9,3.6);

\draw[black, line width=1] (1.7,3.25) -- (3.2,3.25);
\draw[red, photon, line width=1] (1.7,1.05) -- (3.2,1.05);
\draw[red, photon, line width=1] (1.7,-1.15) -- (3.2,-1.15);

 \draw[black, line width=1] (1.35,0.7) -- (1.35,-0.8);
 \draw[black, line width=1] (3.55,0.7) -- (3.55,-0.8);
 \draw[red, photon, line width=1] (3.55,2.9) -- (3.55,1.4);

\filldraw (-0.1,3.25) node {$(A_1,A_2,A_3)$};

\filldraw (-0.3,1.65) node {$(A_4+5A_5,$};
\filldraw (-0.31,1.25) node {$B_3+B_4,$};
\filldraw (-0.1,0.85) node {$A_4+6A_5\,+$};
\filldraw(-0.19,0.45) node {$B_3+2B_4)$};

\filldraw (-0.4,-0.55) node {$(A_6+A_7,$};
\filldraw (-0.22,-0.95) node {$B_3+3B_4,$};
\filldraw (-0.12,-1.35) node {$A_6+2A_7\,+$};
\filldraw (-0.21,-1.75) node {$B_3+4B_4)$};

\filldraw (4.9,3.65) node {$(A_4+A_5,$};
\filldraw (4.97,3.25) node {$B_1+B_2,$};
\filldraw (5.1,2.85) node {$B_1+2B_2)$};

\filldraw (4.9,1.65) node {$(A_4+3A_5,$};
\filldraw (4.88,1.25) node {$B_5+B_6,$};
\filldraw (5.1,0.85) node {$A_4+4A_5\,+$};
\filldraw (5,0.45) node {$B_5+2B_6)$};

\filldraw (4.9,-0.45) node {$(A_6+3A_7,$};
\filldraw (4.975,-0.95) node {$B_5+3B_6,$};
\filldraw (5.1,-1.35) node {$A_6+4A_7\,+$};
\filldraw (5,-1.75) node {$B_5+4B_6)$};

\draw[violet,line width = 1,->] (3.55,2.3) -- (2.65,2.3);
\draw[violet,line width = 1,->] (2.35,1.05) -- (2.35,2.1);

\filldraw (3.1,2.5) node {$t$};
\filldraw (2.5,1.5) node {$t$};

\filldraw (2.35,2.3) node {$W_2$};

\draw[violet,line width = 1,->] (3.55,0.1) -- (2.65,0.1);
\draw[violet,line width = 1,->] (1.35,0.1) -- (2.05,0.1);

\filldraw (3.1,0.3) node {$t$};
\filldraw (1.7,0.3) node {$t$};

\filldraw (2.35,0.1) node {$W_1$};

\end{tikzpicture}
}
\hspace{0in}
\subfigure[]{ 
\begin{tikzpicture}

\filldraw (3.55,3.25) node {$V_1$};
\filldraw (1.35,1.05) node {$V_2$};
\filldraw (3.55,1.05) node {$V_3$};
\filldraw (1.35,-1.15) node {$V_4$};
\filldraw (3.55,-1.15) node {$V_5$};


\draw[black, line width=1] (1,-1.5) -- (1,-0.8);
\draw[black, line width=1] (1,-1.5) -- (1.7,-1.5);
\draw[black, line width=1] (1,-0.8) -- (1.7,-0.8);
\draw[black, line width=1] (1.7,-1.5) -- (1.7,-0.8);

\draw[black, line width=1] (3.2,-1.5) -- (3.2,-0.8);
\draw[black, line width=1] (3.2,-1.5) -- (3.9,-1.5);
\draw[black, line width=1] (3.2,-0.8) -- (3.9,-0.8);
\draw[black, line width=1] (3.9,-1.5) -- (3.9,-0.8);

\draw[black, line width=1] (1,0.7) -- (1,1.4);
\draw[black, line width=1] (1,0.7) -- (1.7,0.7);
\draw[black, line width=1] (1,1.4) -- (1.7,1.4);
\draw[black, line width=1] (1.7,0.7) -- (1.7,1.4);

\draw[black, line width=1] (3.2,0.7) -- (3.2,1.4);
\draw[black, line width=1] (3.2,0.7) -- (3.9,0.7);
\draw[black, line width=1] (3.2,1.4) -- (3.9,1.4);
\draw[black, line width=1] (3.9,0.7) -- (3.9,1.4);


\draw[black, line width=1] (3.2,2.9) -- (3.2,3.6);
\draw[black, line width=1] (3.2,2.9) -- (3.9,2.9);
\draw[black, line width=1] (3.2,3.6) -- (3.9,3.6);
\draw[black, line width=1] (3.9,2.9) -- (3.9,3.6);

\draw[black, line width=1] (1.7,1.05) -- (3.55,2.9);
\draw[red, photon, line width=1] (1.7,1.05) -- (3.2,1.05);
\draw[red, photon, line width=1] (1.7,-1.15) -- (3.2,-1.15);

 \draw[red, photon, line width=1] (1.35,0.7) -- (1.35,-0.8);
 \draw[black, line width=1] (3.55,0.7) -- (3.55,-0.8);
 \draw[black, line width=1] (3.55,2.9) -- (3.55,1.4);



\filldraw (5,3.25) node {$(A_1,A_2,A_3)$};
\filldraw (-0.2,-1.15) node {$(B_1,B_2,B_3)$};

\filldraw (-0.2,1.45) node {$(A_4+A_5,$};
\filldraw (-0.12,1.05) node {$B_4+B_5,$};
\filldraw (-0,0.65) node {$B_4+2B_5)$};

\filldraw (5,1.65) node {$(A_4+3A_5,$};
\filldraw (4.98,1.25) node {$B_6+B_7,$};
\filldraw (5.2,0.85) node {$A_4+4A_5\,+$};
\filldraw (5.1,0.45) node {$B_6+2B_7)$};

\filldraw (5,-0.75) node {$(A_6+A_7,$};
\filldraw (5.18,-1.15) node {$A_6+2A_7,$};
\filldraw (5.2,-1.55) node {$B_6+3B_7)$};

\draw[violet,line width = 1,->] (1.35,0.1) -- (2.05,0.1);

\filldraw (1.7,0.3) node {$t$};

\draw[violet,line width = 1,->] (2.35,-1.15) -- (2.35,-0.1);

\filldraw (2.5,-0.6) node {$t$};

\filldraw (2.35,0.1) node {$W_2$};

\draw[violet,line width = 1,->] (2.35,1.7) -- (2.35,2.3);

\filldraw (2.2,1.9) node {$t$};

\filldraw (2.35,2.55) node {$W_1$};

\end{tikzpicture}
 }
\end{center}
\vspace{-0.2in}
  \caption{\small Two example graphs $G \in \mathcal{G}_{C\geq 4/3}$ and code constructions for 
  rate $4/3$. $W_1 = (a_1, a_2, a_3,a_4), W_2 = (b_1, b_2, b_3,b_4)$ and each $A_i\backslash B_j$ is a generic linear combination of $(a_1,\cdots,a_4)\backslash (b_1,\cdots,b_4)$. 
}
  \label{fig:sufficient}
\end{figure}

We are targeting at rate $L_w/L_v = 4/3$ so that any pair of nodes connected by an edge contain $2L_v = 3L_w/2$ bits. Except from $L_w$ bits from the desired source, at most we can tolerate $2L_v - L_w = L_w/2$ undesired bits (i.e., interference). Then the key is to guarantee for any $W_k$-edge, $k \in \{1,2\}$, the interference from $W_{3-k}$ has at most half source size. That is, $W_{3-k}$ symbols shall be assigned according to $W_k$-edges ($W_k$-components). When there is no internal edge (or residing path contains $1$-color nodes), such interference based assignment automatically ensures the independence (thus decodability) of desired source symbols. We now come back to the examples in Fig.~\ref{fig:sufficient} to see how to implement the above code design idea.

Consider Example $(a)$ first and Example $(b)$ will follow similarly.
We set $L_w/\log_2 p = 4$ so that $W_1 = (a_1, a_2, a_3, a_4)$ and $W_2 = (b_1, b_2, b_3, b_4)$, where each symbol is from a sufficiently large finite field $\mathbb{F}_p$ (the exact field size will be given in the general proof in Section \ref{sec:sufficient}). To achieve rate $R = L_w/L_v = 4/3$, we set $L_v = 3 \log_2 p$, i.e., each $V_n$ contains three symbols from the same field. We generate a number of generic linear combinations of $(a_1,\cdots, a_4) \backslash (b_1,\cdots, b_4)$ and denote them as $(A_1, A_2, \cdots) \backslash (B_1, B_2, \cdots)$. For now, it suffices to view each $A_i \backslash B_j$ as a random linear combination of symbols from $W_1 \backslash W_2$ and if we can collect four linearly independent combinations of $A_i \backslash B_j$, then we can recover $W_1 \backslash W_2$. The detailed randomized construction is again deferred to the general proof. Each one of the three symbols in $V_n$ will be a linear combination of some $A_i$ and $B_j$ symbols. We first assign the $A_i$ 
and $B_j$ symbols 
in each $V_n$ and then linearly combine them to produce the final three symbols in $V_n$.

\vspace{0.1in}
\tikzset{
    photon/.style={decorate, decoration={snake,
      amplitude = 0.5mm,
      segment length = 3mm}, draw=red},
    electron/.style={draw=blue, postaction={decorate},
        decoration={markings,mark=at position .55 with {\arrow[draw=blue]{>}}}},
    gluon/.style={decorate, draw=blue,
        decoration={coil,amplitude=1.5pt, segment length=4pt}} 
}
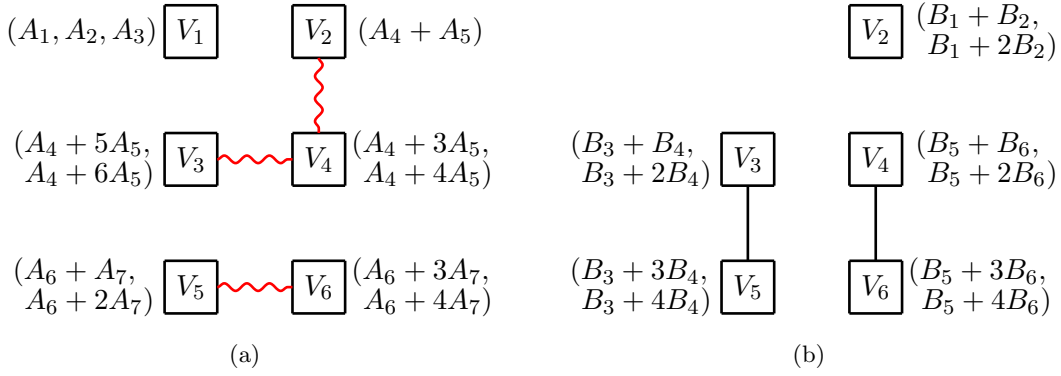
\begin{figure}[H]
\begin{center}
\subfigure[]{
\begin{tikzpicture}

\filldraw (1.35,2.25) node {$V_1$};
\filldraw (3.05,2.25) node {$V_2$};

\filldraw (1.35,0.55) node {$V_3$};
\filldraw (3.05,0.55) node {$V_4$};

\filldraw (1.35,-1.15) node {$V_5$};
\filldraw (3.05,-1.15) node {$V_6$};

\draw[black, line width=1] (1,-1.5) -- (1,-0.8);
\draw[black, line width=1] (1,-1.5) -- (1.7,-1.5);
\draw[black, line width=1] (1,-0.8) -- (1.7,-0.8);
\draw[black, line width=1] (1.7,-1.5) -- (1.7,-0.8);

\draw[black, line width=1] (2.7,-1.5) -- (2.7,-0.8);
\draw[black, line width=1] (2.7,-1.5) -- (3.4,-1.5);
\draw[black, line width=1] (2.7,-0.8) -- (3.4,-0.8);
\draw[black, line width=1] (3.4,-1.5) -- (3.4,-0.8);

\draw[black, line width=1] (1,0.2) -- (1,0.9);
\draw[black, line width=1] (1,0.2) -- (1.7,0.2);
\draw[black, line width=1] (1,0.9) -- (1.7,0.9);
\draw[black, line width=1] (1.7,0.2) -- (1.7,0.9);

\draw[black, line width=1] (2.7,0.2) -- (2.7,0.9);
\draw[black, line width=1] (2.7,0.2) -- (3.4,0.2);
\draw[black, line width=1] (2.7,0.9) -- (3.4,0.9);
\draw[black, line width=1] (3.4,0.2) -- (3.4,0.9);

\draw[black, line width=1] (1,1.9) -- (1,2.6);
\draw[black, line width=1] (1,1.9) -- (1.7,1.9);
\draw[black, line width=1] (1,2.6) -- (1.7,2.6);
\draw[black, line width=1] (1.7,1.9) -- (1.7,2.6);

\draw[black, line width=1] (2.7,1.9) -- (2.7,2.6);
\draw[black, line width=1] (2.7,1.9) -- (3.4,1.9);
\draw[black, line width=1] (2.7,2.6) -- (3.4,2.6);
\draw[black, line width=1] (3.4,1.9) -- (3.4,2.6);


\draw[red, photon, line width=1] (1.7,0.55) -- (2.7,0.55);
\draw[red, photon, line width=1] (1.7,-1.15) -- (2.7,-1.15);

\draw[red, photon, line width=1] (3.05,1.9) -- (3.05,0.9);

\filldraw (-0.1,2.25) node {$(A_1,A_2,A_3)$};
\filldraw (4.4,2.25) node {$(A_4+A_5)$};

\filldraw (4.4,0.75) node {$(A_4+3A_5,$};
\filldraw (4.5,0.35) node {$A_4+4A_5)$};

\filldraw (-0.1,0.75) node {$(A_4+5A_5,$};
\filldraw (0,0.35) node {$A_4+6A_5)$};

\filldraw (4.4,-0.95) node {$(A_6+3A_7,$};
\filldraw (4.5,-1.35) node {$A_6+4A_7)$};

\filldraw (-0.2,-0.95) node {$(A_6+A_7,$};
\filldraw (0,-1.35) node {$A_6+2A_7)$};



\end{tikzpicture}
}
\hspace{0.1in}
\subfigure[]{ 
\begin{tikzpicture}

\filldraw (3.05,2.25) node {$V_2$};

\filldraw (1.35,0.55) node {$V_3$};
\filldraw (3.05,0.55) node {$V_4$};

\filldraw (1.35,-1.15) node {$V_5$};
\filldraw (3.05,-1.15) node {$V_6$};

\draw[black, line width=1] (1,-1.5) -- (1,-0.8);
\draw[black, line width=1] (1,-1.5) -- (1.7,-1.5);
\draw[black, line width=1] (1,-0.8) -- (1.7,-0.8);
\draw[black, line width=1] (1.7,-1.5) -- (1.7,-0.8);

\draw[black, line width=1] (2.7,-1.5) -- (2.7,-0.8);
\draw[black, line width=1] (2.7,-1.5) -- (3.4,-1.5);
\draw[black, line width=1] (2.7,-0.8) -- (3.4,-0.8);
\draw[black, line width=1] (3.4,-1.5) -- (3.4,-0.8);

\draw[black, line width=1] (1,0.2) -- (1,0.9);
\draw[black, line width=1] (1,0.2) -- (1.7,0.2);
\draw[black, line width=1] (1,0.9) -- (1.7,0.9);
\draw[black, line width=1] (1.7,0.2) -- (1.7,0.9);

\draw[black, line width=1] (2.7,0.2) -- (2.7,0.9);
\draw[black, line width=1] (2.7,0.2) -- (3.4,0.2);
\draw[black, line width=1] (2.7,0.9) -- (3.4,0.9);
\draw[black, line width=1] (3.4,0.2) -- (3.4,0.9);


\draw[black, line width=1] (2.7,1.9) -- (2.7,2.6);
\draw[black, line width=1] (2.7,1.9) -- (3.4,1.9);
\draw[black, line width=1] (2.7,2.6) -- (3.4,2.6);
\draw[black, line width=1] (3.4,1.9) -- (3.4,2.6);



\draw[black, line width=1] (1.35,0.2) -- (1.35,-0.8);
\draw[black, line width=1] (3.05,0.2) -- (3.05,-0.8);

\filldraw (4.4,2.45) node {$(B_1+B_2,$};
\filldraw (4.6,2.05) node {$B_1+2B_2)$};

\filldraw (4.4,0.75) node {$(B_5+B_6,$};
\filldraw (4.6,0.35) node {$B_5+2B_6)$};

\filldraw (-0.19,0.75) node {$(B_3+B_4,$};
\filldraw (0,0.35) node {$B_3+2B_4)$};

\filldraw (4.4,-0.95) node {$(B_5+3B_6,$};
\filldraw (4.52,-1.35) node {$B_5+4B_6)$};

\filldraw (-0.1,-0.95) node {$(B_3+3B_4,$};
\filldraw (0,-1.35) node {$B_3+4B_4)$};

\end{tikzpicture}
 }
\end{center}
\vspace{-0.2in}
  \caption{\small (a) $W_2$-component decomposition of $W_1$-connected nodes in Fig.~\ref{fig:sufficient}.(a), according to which $A_i$ symbols are assigned.  (b) $W_1$-component decomposition of $W_2$-connected nodes in Fig.~\ref{fig:sufficient}.(a), according to which $B_j$ symbols are assigned.
  }
  \label{fig:sufficient_component}
\end{figure}

Consider nodes that are connected to $W_1$-edges so that some $A_i$ symbols need to be assigned, i.e., all nodes $V_1,\cdots, V_6$. The $1$-color nodes are trivial (i.e., $V_1$), and we just assign three distinct $A_i$ symbols. Next, consider the remaining $2$-color nodes $V_2, \cdots, V_6$ for which the $A_i$ symbols are assigned according to $W_2$-components (see Fig.~\ref{fig:sufficient_component}.(a)). $V_2, \cdots, V_6$ form two $W_2$-components - one consists of $V_2, V_3, V_4$ and the other consists of $V_5, V_6$. For each $W_2$-component, we assign generic linear combinations of the {\em same} $2 = \frac{1}{2}L_w/\log_2 p$ $A_i$ symbols (say $A_{i_1}, A_{i_2}$) so that the interference dimension is limited to two. Further, a normal $2$-color node and a $W_2$-special $2$-color node will get two generic linear combinations of $(A_{i_1}, A_{i_2})$ and a $W_1$-special $2$-color node will get one generic linear combination of $(A_{i_1}, A_{i_2})$. For example, consider $W_2$-component with nodes $V_2, V_3, V_4$, where the $A_i$ symbols appeared are limited to $A_4, A_5$; $V_2$, as a $W_1$-special $2$-color node, gets one combination $A_4+A_5$ and $V_3, V_4$, as normal $2$-color nodes, each gets two generic combinations (e.g., $V_3$ gets $A_4+5A_5, A_4+6A_5$). The other $W_2$-component with nodes $V_5, V_6$ is assigned similarly - the $A_i$ symbols are limited to $A_6, A_7$.

The assignment for nodes connected to $W_2$-edges is exactly the same (see Fig.~\ref{fig:sufficient_component}.(b)). Nodes $V_2, \cdots, V_6$ are connected to $W_2$-edges and they are all $2$-color. The $B_j$ symbols are assigned according to $W_1$-components, i.e., $V_2$ (as a single-node component) gets generic linear combinations of $B_1, B_2$; $V_3, V_5$ form a $W_1$-component and get generic linear combinations of $B_3, B_4$; $V_4, V_6$ form a $W_1$-component and the $B_j$ symbols are limited to $B_5, B_6$.

The last step is to combine the $A_i, B_j$ symbols so that each $V_n$ has only three symbols. This step is simple, if a node gets at most three $A_i, B_j$ symbols, then just set them as $V_n$ (e.g., $V_1, V_2$); otherwise the node must be normal $2$-color, which gets two generic combinations of $A_i$ and two generic combinations of $B_j$ and we just add one arbitrary combination (say the last) of $A_i$ and $B_j$ together to reduce the total number of symbols to three (e.g., $V_3, V_4, V_5, V_6$).

Finally, let us verify why the decoding constraints (\ref{dec}) are satisfied. An edge that contains $1$-color node is straightforward, e.g., from $W_1$-edge $\{V_1, V_2\}$, we have $A_1, A_2, A_3, A_4 + A_5$, so as long as the $A_i$ combinations are generic we can recover $W_1 = (a_1, \cdots, a_4)$. For edges that connect two $2$-color nodes (e.g., $W_2$-edge $\{V_3, V_4\}$),  we have 1) the interference dimension is limited to two as our assignment is based on components of interfering sources (e.g., we may decode $A_4, A_5$ and remove them, leaving us with only $B_j$ symbols); 2) the four symbols from the desired source have full rank (e.g., $B_3, B_4, B_5, B_6$ are generic combinations) so that we can recover the desired source symbol. Note that because there is no internal edge, for any $W_k$-edge, the two nodes obtain distinct desired $W_k$ symbols, e.g., for $W_2$-edge $\{V_3, V_4\}$, $V_3$ is assigned $B_3, B_4$ symbols and $V_4$ is assigned $B_5, B_6$ symbols as $V_3, V_4$ belong to distinct $W_1$-components (refer to Fig.~\ref{fig:sufficient_component}.(b). If $V_3, V_4$ belong to the same $W_1$-component, then the $W_2$-edge $\{V_3, V_4\}$ will be internal).

The code construction for Example $(b)$ in Fig.~\ref{fig:sufficient} follows from the same procedure as that of Example $(a)$. That is, first consider $1$-color nodes and assign generic combinations (e.g., $V_1, V_4$); for remaining $2$-color nodes, assign $W_k$ symbols according to $W_{3-k}$-components (e.g., the $W_1$ space of the $W_2$-edge $\{V_2, V_3\}$ is spanned by $A_4, A_5$, and the $W_2$ space of the $W_1$-edge $\{V_3, V_5\}$ is spanned by $B_6, B_7$); finally combine the four symbols to three for normal $2$-color nodes (e.g., $V_3$). The decoding constraints (\ref{dec}) are easily verified as the interference dimension is strictly controlled and desired source symbols are sufficiently generic because after removing $1$-color nodes, there no longer exist internal edges.

\subsubsection{Necessary Condition: Residing Path and Special $2$-Color Node}
The sufficient condition of the achievability of rate $4/3$ in Theorem \ref{thm:sufficient} requires the absence of internal edges or the presence of $1$-color node in residing paths. Considering the complementary cases, we identify a crucial graphic structure for the unachievability of rate $4/3$ - the presence of at most one special $2$-color node in a residing path. This result is stated in the following theorem.

\begin{theorem}\label{thm:necessary} [Necessary Condition of $\mathcal{G}_{C=4/3}$]
With $K=2$ source symbols, a graph $G \in \mathcal{G}_{C < 4/3}$ if $G$ has a residing path which contains no $1$-color node and at most one special $2$-color node.
\end{theorem}

The proof of Theorem \ref{thm:necessary} is presented in Section \ref{sec:necessary}. To illustrate the idea, an example is shown in Fig.~\ref{fig:necessary}, where the internal $W_2$-edge $\{V_1, V_2\}$ resides in the $W_1$-path $(\{V_1, V_3\}, \{V_3, V_4\}, \{V_4, V_2\})$ and this residing path contains only one special $2$-color node $V_3$ and no $1$-color node. So the condition of Theorem \ref{thm:necessary} is satisfied and rate 4/3 cannot be achieved.
To see why, we next give an intuitive explanation 
by contradiction.

\vspace{0.1in}
\tikzset{
    photon/.style={decorate, decoration={snake,
      amplitude = 0.5mm,
      segment length = 3mm}, draw=red},
    electron/.style={draw=blue, postaction={decorate},
    decoration={markings,mark=at position .55 with {\arrow[draw=blue]{>}}}},
    gluon/.style={decorate, draw=blue,
        decoration={coil,amplitude=1.5pt, segment length=4pt}} 
}
\begin{figure}[H]
\begin{center}
\begin{tikzpicture}

\filldraw (1.35,2.25) node {$V_1$};
\filldraw (3.05,2.25) node {$V_3$};
\draw[blue, line width=1] (3.05,2.25) circle (0.31);

\filldraw (4.75,2.25) node {$V_5$};

\filldraw (1.35,0.55) node {$V_2$};
\filldraw (3.05,0.55) node {$V_4$};
\filldraw (4.75,0.55) node {$V_6$};
\filldraw (6.45,0.55) node {$V_7$};

 \draw[black, line width=1] (1,0.2) -- (1,0.9);
 \draw[black, line width=1] (1,0.2) -- (1.7,0.2);
 \draw[black, line width=1] (1,0.9) -- (1.7,0.9);
 \draw[black, line width=1] (1.7,0.2) -- (1.7,0.9);

\draw[black, line width=1] (2.7,0.2) -- (2.7,0.9);
\draw[black, line width=1] (2.7,0.2) -- (3.4,0.2);
\draw[black, line width=1] (2.7,0.9) -- (3.4,0.9);
\draw[black, line width=1] (3.4,0.2) -- (3.4,0.9);

\draw[black, line width=1] (1,1.9) -- (1,2.6);
\draw[black, line width=1] (1,1.9) -- (1.7,1.9);
\draw[black, line width=1] (1,2.6) -- (1.7,2.6);
\draw[black, line width=1] (1.7,1.9) -- (1.7,2.6);

\draw[black, line width=1] (2.7,1.9) -- (2.7,2.6);
\draw[black, line width=1] (2.7,1.9) -- (3.4,1.9);
\draw[black, line width=1] (2.7,2.6) -- (3.4,2.6);
\draw[black, line width=1] (3.4,1.9) -- (3.4,2.6);

\draw[black, line width=1] (4.4,0.2) -- (4.4,0.9);
\draw[black, line width=1] (4.4,0.2) -- (5.1,0.2);
\draw[black, line width=1] (4.4,0.9) -- (5.1,0.9);
\draw[black, line width=1] (5.1,0.2) -- (5.1,0.9);

\draw[black, line width=1] (6.1,0.2) -- (6.1,0.9);
\draw[black, line width=1] (6.1,0.2) -- (6.8,0.2);
\draw[black, line width=1] (6.1,0.9) -- (6.8,0.9);
\draw[black, line width=1] (6.8,0.2) -- (6.8,0.9);

 \draw[black, line width=1] (4.4,1.9) -- (4.4,2.6);
\draw[black, line width=1] (4.4,1.9) -- (5.1,1.9);
\draw[black, line width=1] (4.4,2.6) -- (5.1,2.6);
 \draw[black, line width=1] (5.1,1.9) -- (5.1,2.6);

\draw[black, line width=1] (1.7,2.25) -- (2.7,2.25);
\draw[red, photon,  line width=2] (1.35,1.9) -- (1.35,0.9);
\draw[black, line width=1] (1.7,0.55) -- (2.7,0.55);

\draw[black, line width=1] (3.05,1.9) -- (3.05,0.9);

\draw[red, photon, line width=1] (3.4,2.25) -- (4.4,2.25);
\draw[red, photon, line width=1] (3.4,0.55) -- (4.4,0.55);
\draw[black, line width=1] (5.1,0.55) -- (6.1,0.55);

\draw[black, line width=1] (8.2,1.55) -- (7.4,1.55);
\draw[red, photon, line width=1] (8.2,0.55) -- (7.4,0.55);

\filldraw (8.6,2.35) node {$t$};
\filldraw (8.6,1.55) node {$W_1$};
\filldraw (8.6,0.55) node {$W_2$};

\end{tikzpicture}

\end{center}
\vspace{-0.1in}
  \caption{\small An example graph $G \in \mathcal{G}_{C < 4/3}$ with internal edge $\{V_1, V_2\}$ highlighted and the only special $2$-color node $V_3$ in its residing path highlighted.
  }
  \label{fig:necessary}
\end{figure}
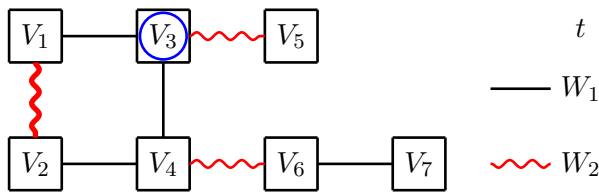

Suppose rate $4/3$ is achievable, i.e., $L_w/L_v = 4/3$. Then we can show that for any $2$-color node (e.g., $V_3$), it must contain at least $L_w/4$ bits of information about each of $W_1$ and $W_2$ (captured through conditional entropy. See Lemma \ref{lemma:2color} in Section \ref{sec:necessary}). This is because the connecting node can provide at most $L_v = 3L_w/4$ bits of information about the desired source symbol (e.g., $V_5$ can contribute $L_v = 3L_w/4$ bits on $W_2$ at most and the remaining $L_w - L_v = L_w/4$ bits must come from $V_3$). Further, if the $2$-color node is normal (e.g., $V_4$), it must contain exactly $L_w/2$ bits of information about each of $W_1$ and $W_2$ (see Lemma \ref{lemma:normal}). The reason is that for two connecting $2$-color nodes, the amount of interference allowed is at most $2L_v - L_w = L_w/2$ bits and a pair of nodes must contribute $L_w$ bits of information about the desired source symbol (thus $L_w/2$ from each node). For example, consider $W_1$-edge $\{V_2, V_4\}$, where from an interference view, $V_2$ can contain at most $L_w/2$ bits on $W_2$; from the desired source view, $V_2$ must also contribute at least $L_w/2$ bits on $W_2$ because of the $W_2$-edge $\{V_1,V_2\}$.

We now consider the propagation of interference through the residing $W_1$-path $(\{V_2, V_4\}$, $\{V_4, V_3\}$, $\{V_3, V_1\})$. Start from the normal $2$-color node $V_2$, which contains $L_w/2$ bits on $W_2$ and as a $W_1$-edge can tolerate at most $L_w/2$ bits on $W_2$, then the normal $2$-color node $V_4$ must contain the same $L_w/2$ bits on $W_2$ (see Lemma \ref{lemma:edge}). We are now at $V_4$ and continue the $W_1$-path through edge $\{V_3, V_4\}$, where $V_3$ is special so that $V_3$ contains at least $L_w/4$ bits on $W_2$ and this $L_w/4$ bits are contained in the total $L_w/2$ interference bits in $V_4$. Continue further the $W_1$-path through edge $\{V_3, V_1\}$, where the $L_w/4$ bits on $W_2$ in $V_3$ must be contained in the $L_w/2$ bits on $W_2$ in $V_1$. This in turn means that the $L_w/2$ bits on $W_2$ in $V_1$ must overlap with the $L_w/2$ bits on $W_2$ in $V_2$ (in the $L_w/4$ bits on $W_2$ in $V_3$), thus the internal $W_2$-edge $\{V_1, V_2\}$ cannot contribute $L_w/2 + L_w/2 = L_w$ independent bits for the desired $W_2$ source and we have arrived at a contradiction.

From the above reasoning, we can now illuminate the role of special and normal $2$-color nodes in a residing path. For an internal $W_k$-edge, its residing $W_{3-k}$-path made up of $2$-color nodes must have two normal $2$-color end nodes, each of which contains $L_w/2$ independent bits of information about the desired source $W_k$ (e.g., $V_1, V_2$ about $W_2$). In the residing $W_{3-k}$-path, a normal $2$-color node will keep the interference on $W_k$ to the same $L_w/2$ dimensions (e.g., $V_2, V_4$ have the same $L_w/2$ dimensions about $W_2$ and $V_1,V_3$ have the same $L_w/2$ dimensions about $W_2$) while a special $2$-color node will inherit at least $L_w/4$ interference dimensions on $W_k$ (e.g., $V_3$ gets at least $L_w/4$ dimensions of $W_2$ from $V_3$). Conversely, a special $2$-color node in a residing $W_{3-k}$-path can change at most $L_w/4$ dimensions of the interference on $W_k$ (which is the desired source for the internal $W_k$-edge), so to ensure the independence of the desired source at the internal edge we need at least two special $2$-color nodes in the residing path. This case is exactly our focus in the next section (along this line, we can also see the role of $1$-color node in a residing path, i.e., it completely stops the propagation of interference. See the $1$-color node $V_4$ in the residing $W_2$-path $(\{V_3, V_2\}, \{V_2,V_4\}, \{V_4, V_5\})$ of Fig.~\ref{fig:sufficient_component}.(b), where $V_3, V_5$ can hold independent $W_1$ bits although the $W_1$-edge $\{V_3, V_5\}$ is internal).

\subsubsection{Remaining Cases: Rate $4/3$ May or May Not be Achievable}

Continuing the discussion in the previous paragraph, the cases that are not covered by Theorem \ref{thm:sufficient} and Theorem \ref{thm:necessary} are those where each residing path contains at least two special $2$-color nodes (and no $1$-color node). This setting turns out to be quite intricate and is not fully understood. In the following, we show that here $4/3$ may or may not be achievable, depending on the structure of other parts of the graph.

On the one hand, we show that if there is only one internal edge, then the presence of two special $2$-color nodes in the residing path is sufficient to achieve rate $4/3$. This result is stated in the following theorem.

\begin{theorem}\label{thm:ach}
With $K=2$ source symbols, a graph $G \in \mathcal{G}_{C \geq 4/3}$ if $G$ contains only one internal edge and its every residing path has at least two special $2$-color nodes.
\end{theorem}

\vspace{0.1in}
\tikzset{
    photon/.style={decorate, decoration={snake,
      amplitude = 0.5mm,
      segment length = 3mm}, draw=red},
    electron/.style={draw=blue, postaction={decorate},
    decoration={markings,mark=at position .55 with {\arrow[draw=blue]{>}}}},
    gluon/.style={decorate, draw=blue,
        decoration={coil,amplitude=1.5pt, segment length=4pt}} 
}
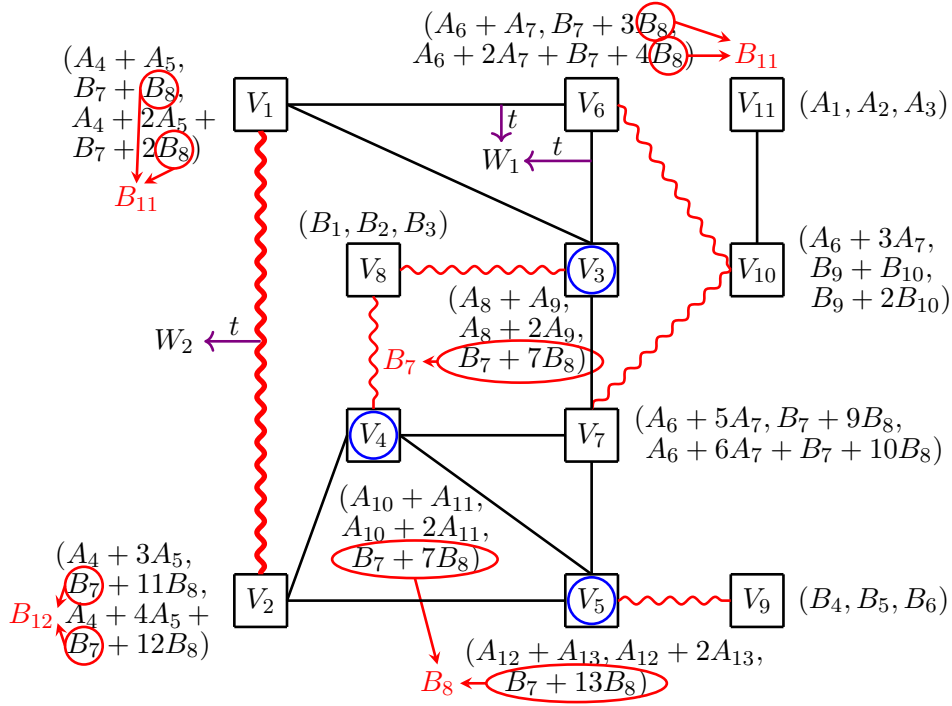
\begin{figure}[H]
\begin{center}
\begin{tikzpicture}
\filldraw (-0.85,5.45) node {$V_1$};
\filldraw (-2.67,6.05) node {$(A_4+A_5,$};
\filldraw (-2.6,5.65) node {$B_7+B_8,$};
\filldraw (-2.4,5.25) node {$A_4+2A_5\,+$};
\filldraw (-2.49,4.85) node {$B_7+2B_8)$};
\draw [red, line width=1](-2.0,4.85) ellipse (0.25 and 0.25 );
\draw [red, line width=1](-2.2,5.65) ellipse (0.25 and 0.25 );
\draw[-stealth] [red, line width=1] (-2.0,4.6) -- (-2.4,4.4);
\draw[-stealth] [red, line width=1] (-2.45,5.65) -- (-2.5,4.4);
\filldraw (-2.5,4.2) node {$\textcolor{red}{B_{11}}$};

\filldraw (-0.85,-1.15) node {$V_2$};
\filldraw (-2.67,-0.55) node {$(A_4+3A_5,$};
\filldraw (-2.49,-0.95) node {$B_7+11B_8,$};
\filldraw (-2.49,-1.35) node {$A_4+4A_5\,+$};
\filldraw (-2.47,-1.75) node {$B_7+12B_8)$};
\draw [red, line width=1](-3.2,-1.75) ellipse (0.25 and 0.25 );
\draw [red, line width=1](-3.2,-0.95) ellipse (0.25 and 0.25 );

\draw[-stealth] [red, line width=1] (-3.45,-1.75) -- (-3.55,-1.45);

\draw[-stealth] [red, line width=1] (-3.45,-0.95) -- (-3.55,-1.25);

\filldraw (-3.9,-1.35) node {$\textcolor{red}{B_{12}}$};

\filldraw (3.55,3.25) node {$V_3$};
\filldraw (2.45,2.85) node {$(A_8+A_9,$};
\filldraw (2.62,2.45) node {$A_8+2A_9,$};
\filldraw (2.63,2.05) node {$B_7+7B_8)$};
\draw[blue, line width=1] (3.55,3.25) circle (0.31);

\draw [red, line width=1](2.6,2.03) ellipse (1.1 and 0.28 );
\draw[-stealth] [red, line width=1] (1.5,2.03) -- (1.3,2.03);
\filldraw (1,2.03) node {$\textcolor{red}{B_7}$};

\filldraw (0.65,1.05) node {$V_4$};
\filldraw (1.15,0.2) node {$(A_{10}+A_{11},$};
\filldraw (1.21,-0.2) node {$A_{10}+2A_{11},$};
\filldraw (1.22,-0.6) node {$B_7+7B_8)$};
\draw[blue, line width=1] (0.65,1.05) circle (0.31);

\draw [red, line width=1](1.2,-0.6) ellipse (1.1 and 0.25 );
\draw[-stealth] [red, line width=1] (1.2,-0.85) -- (1.5,-2);

\filldraw (3.55,-1.15) node {$V_5$};
\filldraw (3.85,-1.85) node {$(A_{12}+A_{13},A_{12}+2A_{13},$};
\filldraw (3.35,-2.25) node {$B_7+13B_8)$};
\draw[blue, line width=1] (3.55,-1.15) circle (0.31);

\draw [red, line width=1](3.35,-2.25) ellipse (1.2 and 0.25 );
\draw[-stealth] [red, line width=1] (2.15,-2.25) -- (1.8,-2.25);
\filldraw (1.5,-2.25) node {$\textcolor{red}{B_8}$};

\filldraw (3.55,5.45) node {$V_6$};
\filldraw (3,6.5) node {$(A_6+A_7,B_7+3B_8,$};
\filldraw (3.05,6.1) node {$A_6+2A_7+B_7+4B_8)$};

\draw [red, line width=1](4.57,6.1) ellipse (0.25 and 0.25 );
\draw [red, line width=1](4.4,6.55) ellipse (0.25 and 0.25 );
\draw[-stealth] [red, line width=1] (4.82,6.1) -- (5.4,6.1);
\draw[-stealth] [red, line width=1] (4.65,6.55) -- (5.4,6.3);
\filldraw (5.75,6.1) node {$\textcolor{red}{B_{11}}$};

\filldraw (3.55,1.05) node {$V_7$};
\filldraw (5.9,1.25) node {$(A_6+5A_7,B_7+9B_8,$};
\filldraw (6.25,0.85) node {$A_6+6A_7+B_7+10B_8)$};

\filldraw (0.65,3.25) node {$V_8$};
\filldraw (0.65,3.85) node {$(B_1,B_2,B_3)$};

\filldraw (5.75,-1.15) node {$V_{9}$};
\filldraw (7.3,-1.15) node {$(B_4,B_5,B_6)$};

\filldraw (5.75,3.25) node {$V_{10}$};
\filldraw (7.2,3.65) node {$(A_6+3A_7,$};
\filldraw (7.27,3.25) node {$B_9+B_{10},$};
\filldraw (7.4,2.85) node {$B_9+2B_{10})$};

\filldraw (5.75,5.45) node {$V_{11}$};
\filldraw (7.3,5.45) node {$(A_1,A_2,A_3)$};

\draw[black, line width=1] (-1.2,-1.5) -- (-1.2,-0.8);
\draw[black, line width=1] (-1.2,-1.5) -- (-0.5,-1.5);
\draw[black, line width=1] (-1.2,-0.8) -- (-0.5,-0.8);
\draw[black, line width=1] (-0.5,-1.5) -- (-0.5,-0.8);

\draw[black, line width=1] (-1.2,5.1) -- (-1.2,5.8);
\draw[black, line width=1] (-1.2,5.1) -- (-0.5,5.1);
\draw[black, line width=1] (-1.2,5.8) -- (-0.5,5.8);
\draw[black, line width=1] (-0.5,5.1) -- (-0.5,5.8);

\draw[black, line width=1] (3.2,-1.5) -- (3.2,-0.8);
\draw[black, line width=1] (3.2,-1.5) -- (3.9,-1.5);
\draw[black, line width=1] (3.2,-0.8) -- (3.9,-0.8);
\draw[black, line width=1] (3.9,-1.5) -- (3.9,-0.8);

\draw[black, line width=1] (0.3,0.7) -- (0.3,1.4);
\draw[black, line width=1] (0.3,0.7) -- (1,0.7);
\draw[black, line width=1] (0.3,1.4) -- (1,1.4);
\draw[black, line width=1] (1,0.7) -- (1,1.4);

\draw[black, line width=1] (3.2,0.7) -- (3.2,1.4);
\draw[black, line width=1] (3.2,0.7) -- (3.9,0.7);
\draw[black, line width=1] (3.2,1.4) -- (3.9,1.4);
\draw[black, line width=1] (3.9,0.7) -- (3.9,1.4);

\draw[black, line width=1] (0.3,2.9) -- (0.3,3.6);
\draw[black, line width=1] (0.3,2.9) -- (1,2.9);
\draw[black, line width=1] (0.3,3.6) -- (1,3.6);
\draw[black, line width=1] (1,2.9) -- (1,3.6);

\draw[black, line width=1] (3.2,2.9) -- (3.2,3.6);
\draw[black, line width=1] (3.2,2.9) -- (3.9,2.9);
\draw[black, line width=1] (3.2,3.6) -- (3.9,3.6);
\draw[black, line width=1] (3.9,2.9) -- (3.9,3.6);

\draw[black, line width=1] (3.2,5.1) -- (3.2,5.8);
\draw[black, line width=1] (3.2,5.1) -- (3.9,5.1);
\draw[black, line width=1] (3.2,5.8) -- (3.9,5.8);
\draw[black, line width=1] (3.9,5.1) -- (3.9,5.8);

\draw[black, line width=1] (5.4,5.1) -- (5.4,5.8);
\draw[black, line width=1] (5.4,5.1) -- (6.1,5.1);
\draw[black, line width=1] (5.4,5.8) -- (6.1,5.8);
\draw[black, line width=1] (6.1,5.1) -- (6.1,5.8);

\draw[black, line width=1] (5.4,2.9) -- (5.4,3.6);
\draw[black, line width=1] (5.4,2.9) -- (6.1,2.9);
\draw[black, line width=1] (5.4,3.6) -- (6.1,3.6);
\draw[black, line width=1] (6.1,2.9) -- (6.1,3.6);

\draw[black, line width=1] (5.4,-1.5) -- (5.4,-0.8);
\draw[black, line width=1] (5.4,-1.5) -- (6.1,-1.5);
\draw[black, line width=1] (5.4,-0.8) -- (6.1,-0.8);
\draw[black, line width=1] (6.1,-1.5) -- (6.1,-0.8);

\draw[red, photon, line width=1] (1,3.25) -- (3.2,3.25);
\draw[red, photon, line width=1] (0.65,2.9) -- (0.65,1.4);
\draw[red, photon, line width=2] (-0.85,5.1) -- (-0.85,-0.8);
\draw[red, photon, line width=1] (3.9,-1.15) -- (5.4,-1.15);
\draw[red, photon, line width=1] (3.55,1.4) -- (5.4,3.25);
\draw[red, photon, line width=1] (3.9,5.45) -- (5.4,3.25);

\draw[black, line width=1] (1,1.05) -- (3.55,-0.8);
\draw[black, line width=1] (0.3,1.05) -- (-0.5,-1.15);
\draw[black, line width=1] (1,1.05) -- (3.2,1.05);
\draw[black, line width=1] (-0.5,-1.15) -- (3.2,-1.15);
\draw[black, line width=1] (-0.5,5.45) -- (3.2,5.45);
\draw[black, line width=1] (-0.5,5.45) -- (3.55,3.6);

\draw[black, line width=1] (3.55,0.7) -- (3.55,-0.8);
\draw[black, line width=1] (3.55,2.9) -- (3.55,1.4);
\draw[black, line width=1] (3.55,5.1) -- (3.55,3.6);
\draw[black, line width=1] (5.75,5.1) -- (5.75,3.6);

 \draw[violet,line width = 1,->] (3.55,4.7) -- (2.65,4.7);
\draw[violet,line width = 1,->] (2.35,5.45) -- (2.35,4.95);

 \filldraw (3.1,4.9) node {$t$};
\filldraw (2.5,5.25) node {$t$};

 \filldraw (2.35,4.7) node {$W_1$};

\draw[violet,line width = 1,->] (-0.85,2.3) -- (-1.6,2.3);

 \filldraw (-1.2,2.5) node {$t$};

 \filldraw (-2,2.3) node {$W_2$};

\end{tikzpicture}

\end{center}
\vspace{-0.1in}
  \caption{\small An example graph $G \in \mathcal{G}_{C\geq 4/3}$ with only one internal edge $\{V_1,V_2\}$ (highlighted). Special $2$-color nodes $V_3, V_4, V_5$ are highlighted and each residing path has at least two of them. 
  The code produced by the assignment of Theorem \ref{thm:sufficient} and updates needed to produce the final code with rate $4/3$ are shown.
  }
  \label{fig:ach}
\end{figure}

The proof of Theorem \ref{thm:ach} is presented in Section \ref{sec:ach}. To illustrate the idea, an example is shown in Fig.~\ref{fig:ach}, where for the only internal edge $\{V_1, V_2\}$, three special $2$-color nodes $V_3, V_4, V_5$ ensure that at least two of them are contained in any residing path. 

To assign the code, we first follow the same procedure in Theorem \ref{thm:sufficient} to assign the coded symbols (see Fig.~\ref{fig:ach}). Because of the presence of the internal $W_2$-edge $\{V_1, V_2\}$ (and absence of $1$-color nodes in the residing paths), the desired $W_2$ symbols are not independent (i.e., only $B_7, B_8$ appears in $V_1, V_2$ while we need four $B_j$ symbols to recover $W_2$). So we need to expand the dimension of the $W_2$ symbols to satisfy the decoding constraint (\ref{dec}) for the internal edge $\{V_1, V_2\}$. This is done by replacing $B_8 \backslash B_7$ in $V_1 \backslash V_2$ with another generic $B_{11} \backslash B_{12}$ symbol (see Fig.~\ref{fig:ach}), but now the interference on $W_2$ in the residing path will not be limited to only two dimensions. A final update is required - starting from $V_2$, we visit each residing path to find its closest special $2$-color nodes, which turn out to be $V_4, V_5$ and remove $B_7$ therein to ensure the interference along this path is limited to $B_8, B_{12}$ two dimensions (see Fig.~\ref{fig:ach}). Repeat the same for $V_1$, i.e., visit each residing path starting from $V_1$, find the closest special $2$-color nodes, which turn out to be $V_3$, and only keep $B_7$ (remove $B_8$) at $V_3$. Such special $2$-color nodes are guaranteed to exist as each residing path has at least two special $2$-color nodes. Also replace $B_8$ by $B_{11}$ for each node visited along the residing paths so that now again the interference dimension is limited to $B_7, B_{11}$ (i.e., $V_6$. See Fig.~\ref{fig:ach}). The update is complete and decoding constraints (\ref{dec}) are all satisfied (refer to Fig.~\ref{fig:ach} for a verification). Indeed, we may see that the role of each special $2$-color node along the residing $W_k$-path is to replace $L_w/4$ dimensions of $W_{3-k}$ so that with two special $2$-color nodes we may have fully independent $L_w/2$ dimensions of $W_{3-k}$ for the two nodes in the internal $W_{3-k}-$edge.

\bigskip

On the other hand, we show that for the graph in Fig.~\ref{fig:not_ach}, rate $4/3$ cannot be achieved even if each residing path contains two special $2$-color nodes. This result is stated in the following theorem.

\vspace{0.1in}
\tikzset{
    photon/.style={decorate, decoration={snake,
      amplitude = 0.5mm,
      segment length = 3mm}, draw=red},
    electron/.style={draw=blue, postaction={decorate},
    decoration={markings,mark=at position .55 with {\arrow[draw=blue]{>}}}},
    gluon/.style={decorate, draw=blue,
        decoration={coil,amplitude=1.5pt, segment length=4pt}} 
}
\begin{figure}[H]
\begin{center}
\begin{tikzpicture}

\filldraw (1.35,2.25) node {$V_1$};
\filldraw (1.35,0.55) node {$V_2$};
\filldraw (8.15,2.25) node {$V_3$};
\filldraw (8.15,0.55) node {$V_4$};

\filldraw (4.75,2.25) node {$V_5$};
\draw[blue, line width=1] (4.75,2.25) circle (0.31);

\filldraw (3.05,0.55) node {$V_6$};
\draw[blue, line width=1] (3.05,0.55) circle (0.31);

\filldraw (6.45,0.55) node {$V_7$};
\draw[blue, line width=1] (6.45,0.55) circle (0.31);

\filldraw (4.75,0.55) node {$V_8$};
\filldraw (2.65,1.55) node {$V_9$};

\filldraw (1.35,-1.15) node {$V_{10}$};
\filldraw (3.05,-1.15) node {$V_{11}$};
\filldraw (4.75,-1.15) node {$V_{12}$};
\filldraw (6.45,-1.15) node {$V_{13}$};

\draw[black, line width=1] (1,-1.5) -- (1,-0.8);
\draw[black, line width=1] (1,-1.5) -- (1.7,-1.5);
\draw[black, line width=1] (1,-0.8) -- (1.7,-0.8);
\draw[black, line width=1] (1.7,-1.5) -- (1.7,-0.8);

\draw[black, line width=1] (1,0.2) -- (1,0.9);
\draw[black, line width=1] (1,0.2) -- (1.7,0.2);
\draw[black, line width=1] (1,0.9) -- (1.7,0.9);
\draw[black, line width=1] (1.7,0.2) -- (1.7,0.9);

\draw[black, line width=1] (2.7,0.2) -- (2.7,0.9);
\draw[black, line width=1] (2.7,0.2) -- (3.4,0.2);
\draw[black, line width=1] (2.7,0.9) -- (3.4,0.9);
\draw[black, line width=1] (3.4,0.2) -- (3.4,0.9);

\draw[black, line width=1] (2.3,1.2) -- (2.3,1.9);
\draw[black, line width=1] (2.3,1.2) -- (3,1.2);
\draw[black, line width=1] (2.3,1.9) -- (3,1.9);
\draw[black, line width=1] (3,1.2) -- (3,1.9);

\draw[red, photon, line width=1] (3,1.55) -- (4.4,2.25);

\draw[black, line width=1] (1,1.9) -- (1,2.6);
\draw[black, line width=1] (1,1.9) -- (1.7,1.9);
\draw[black, line width=1] (1,2.6) -- (1.7,2.6);
\draw[black, line width=1] (1.7,1.9) -- (1.7,2.6);

\draw[black, line width=1] (4.4,-1.5) -- (4.4,-0.8);
\draw[black, line width=1] (4.4,-1.5) -- (5.1,-1.5);
\draw[black, line width=1] (4.4,-0.8) -- (5.1,-0.8);
\draw[black, line width=1] (5.1,-1.5) -- (5.1,-0.8);

\draw[black, line width=1] (4.4,0.2) -- (4.4,0.9);
\draw[black, line width=1] (4.4,0.2) -- (5.1,0.2);
\draw[black, line width=1] (4.4,0.9) -- (5.1,0.9);
\draw[black, line width=1] (5.1,0.2) -- (5.1,0.9);

\draw[black, line width=1] (6.1,0.2) -- (6.1,0.9);
\draw[black, line width=1] (6.1,0.2) -- (6.8,0.2);
\draw[black, line width=1] (6.1,0.9) -- (6.8,0.9);
\draw[black, line width=1] (6.8,0.2) -- (6.8,0.9);

\draw[black, line width=1] (4.4,1.9) -- (4.4,2.6);
\draw[black, line width=1] (4.4,1.9) -- (5.1,1.9);
\draw[black, line width=1] (4.4,2.6) -- (5.1,2.6);
\draw[black, line width=1] (5.1,1.9) -- (5.1,2.6);

\draw[black, line width=1] (7.8,1.9) -- (7.8,2.6);
\draw[black, line width=1] (7.8,1.9) -- (8.5,1.9);
\draw[black, line width=1] (7.8,2.6) -- (8.5,2.6);
\draw[black, line width=1] (8.5,1.9) -- (8.5,2.6);

\draw[black, line width=1] (7.8,0.2) -- (7.8,0.9);
\draw[black, line width=1] (7.8,0.2) -- (8.5,0.2);
\draw[black, line width=1] (7.8,0.9) -- (8.5,0.9);
\draw[black, line width=1] (8.5,0.2) -- (8.5,0.9);

\draw[black, line width=1] (6.1,-1.5) -- (6.1,-0.8);
\draw[black, line width=1] (6.1,-1.5) -- (6.8,-1.5);
\draw[black, line width=1] (6.1,-0.8) -- (6.8,-0.8);
\draw[black, line width=1] (6.8,-1.5) -- (6.8,-0.8);

\draw[black, line width=1] (2.7,-1.5) -- (2.7,-0.8);
\draw[black, line width=1] (2.7,-1.5) -- (3.4,-1.5);
\draw[black, line width=1] (2.7,-0.8) -- (3.4,-0.8);
\draw[black, line width=1] (3.4,-1.5) -- (3.4,-0.8);

\draw[red, photon, line width=2] (1.35,1.9) -- (1.35,0.9);
\draw[red, photon, line width=2] (8.15,1.9) -- (8.15,0.9);
\draw[red, photon, line width=1] (4.75,0.2) -- (4.75,-0.8);
\draw[red, photon, line width=1] (6.45,0.2) -- (6.45,-0.8);
\draw[red, photon, line width=1] (1.7,-1.15) -- (2.7,0.55);

\draw[black, line width=1] (4.75,1.9) -- (4.75,0.9);
\draw[black, line width=1] (1.7,2.25) -- (4.4,2.25);
\draw[black, line width=1] (5.1,2.25) -- (7.8,2.25);

\draw[black, line width=1] (3.05,0.9) -- (4.66,1.36);
\draw[black, line width=1] (4.84,1.43) -- (7.8,2.25);

\draw [black,thick,domain=10:190] plot ({4.75+0.1*cos(\x)}, {1.4+0.15*sin(\x)});

\draw[black, line width=1] (1.7,0.55) -- (2.7,0.55);
\draw[black, line width=1] (3.4,0.55) -- (4.4,0.55);
\draw[black, line width=1] (5.1,0.55) -- (6.1,0.55);
\draw[black, line width=1] (6.8,0.55) -- (7.8,0.55);

\draw[black, line width=1] (3.4,-1.15) -- (4.4,-1.15);

\draw[black, line width=1] (9.9,0.5) -- (9.1,0.5);
\draw[red, photon, line width=1] (9.9,-0.5) -- (9.1,-0.5);

\filldraw (10.3,1.3) node {$t$};

\filldraw (10.3,0.5) node {$W_1$};
\filldraw (10.3,-0.5) node {$W_2$};

\end{tikzpicture}

\end{center}
\vspace{-0.1in}
  \caption{\small A graph $G \in \mathcal{G}_{C<4/3}$ albeit each residing path has two special $2$-color nodes. $G$ contains two internal edges $\{V_1, V_2\}$, $\{V_3, V_4\}$ and three special $2$-color nodes $V_5, V_6, V_7$ in residing paths.}
\label{fig:not_ach}
\end{figure}
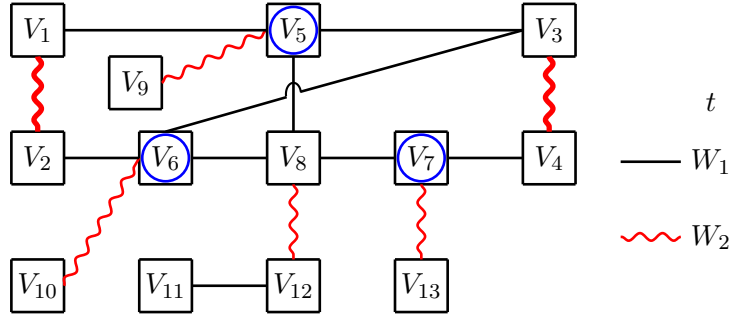

\begin{theorem}\label{thm:not_ach}
The storage code capacity of the graph $G$ in Fig.~\ref{fig:not_ach} is strictly smaller than $4/3$.
\end{theorem}

The proof of Theorem \ref{thm:not_ach} is presented in Section \ref{sec:not_ach}. An intuitive explanation, which builds upon and generalizes the converse arguments in Theorem \ref{thm:necessary}, on the unachievablity of rate $4/3$ is given here. Suppose rate $4/3$ can be achieved. Consider the internal $W_2$-edge $\{V_1, V_2\}$, where $V_1, V_2$ are normal $2$-color and each must contain independent $L_w/2$ bits of information about $W_2$. Due to the $W_1$-edges $\{V_1, V_5\}$ and $\{V_2, V_6\}$, the special $2$-color node $V_5$ must inherit $L_w/4$ bits on $W_2$ from $V_1$ (because the total amount of interference about $W_2$ in any $W_1$-edge cannot exceed $L_w/2$ bits) and the special $2$-color node $V_6$ must inherit $L_w/4$ bits on $W_2$ from $V_2$. 
Note now that $V_3$ is connected to both $V_5, V_6$ with $W_1$-edges, so the $L_w/2$ bits of information about $W_2$ in the normal $2$-color node $V_3$ must contain the $L_w/4$ bits of information about $W_2$ in $V_5$ and $V_6$. Further, $V_5$ and $V_6$ contain independent information about $W_2$. So the $L_w/2$ bits on $W_2$ in $V_3$ is exactly the union of the $L_w/4$ bits on $W_2$ in $V_5$ and $V_6$. From the same reasoning, as $V_8$ is connected to both $V_5$ and $V_6$ with $W_1$-edges, $V_8$ contains exactly the same $L_w/2$ bits of information about $W_2$ as $V_3$. This will cause a contradiction because $(\{V_8, V_7\}, \{V_7, V_4\})$ belongs to a residing path and $V_7$ is a special $2$-color node, so $V_4$ must share $L_w/4$ bits on $W_2$ with $V_8$ (thus $V_3$), which contradicts the fact that $\{V_3,V_4\}$ is an internal $W_2$-edge, i.e., $V_3, V_4$ must contain independent $L_w/2$ bits of information about $W_2$.

\subsection{Extremal Graphs with Capacity $4/3$: $\mathcal{G}_{C = 4/3}$ with $K>2$ Source Symbols}

We now generalize the results on $\mathcal{G}_{C = 4/3}$ from $K=2$ source symbols to $K>2$ source symbols. Let us start from necessary conditions, which include some new graphic structures with more than $2$ sources that place rate constraints, and state the result in the following theorem.

\begin{theorem}\label{thm:necessary2} [Necessary Condition of $\mathcal{G}_{C=4/3}$]
A graph $G \in \mathcal{G}_{C < 4/3}$ if $G$ contains 1) an $M$-color node, where $M \geq 4$, or 2) a $3$-color code that is connected to an $M$-color code, where $M \geq 2$, or 3) a normal $2$-color node $V$ that is connected to a $2$-color node whose connected edges are associated with a different set of source symbols from that connected to $V$. 
\end{theorem}

\vspace{0.1in}
\tikzset{
    photon/.style={decorate, decoration={snake,
      amplitude = 0.5mm,
      segment length = 3mm}, draw=red},
    electron/.style={draw=blue, postaction={decorate},
    decoration={markings,mark=at position .55 with {\arrow[draw=blue]{>}}}},
    gluon/.style={decorate, draw=blue,
        decoration={coil,amplitude=1.5pt, segment length=4pt}} 
}
\begin{figure}[H]
\begin{center}
\begin{tikzpicture}

\filldraw (1.35,2.25) node {$V_1$};
\filldraw (3.05,2.25) node {$V_2$};
\filldraw (4.75,2.25) node {$V_3$};

\filldraw (1.35,0.55) node {$V_4$};
\filldraw (3.05,0.55) node {$V_5$};

\draw[black, line width=1] (1,0.2) -- (1,0.9);
\draw[black, line width=1] (1,0.2) -- (1.7,0.2);
\draw[black, line width=1] (1,0.9) -- (1.7,0.9);
\draw[black, line width=1] (1.7,0.2) -- (1.7,0.9);

\draw[black, line width=1] (2.7,0.2) -- (2.7,0.9);
\draw[black, line width=1] (2.7,0.2) -- (3.4,0.2);
\draw[black, line width=1] (2.7,0.9) -- (3.4,0.9);
\draw[black, line width=1] (3.4,0.2) -- (3.4,0.9);

\draw[black, line width=1] (1,1.9) -- (1,2.6);
\draw[black, line width=1] (1,1.9) -- (1.7,1.9);
\draw[black, line width=1] (1,2.6) -- (1.7,2.6);
\draw[black, line width=1] (1.7,1.9) -- (1.7,2.6);

\draw[black, line width=1] (2.7,1.9) -- (2.7,2.6);
\draw[black, line width=1] (2.7,1.9) -- (3.4,1.9);
\draw[black, line width=1] (2.7,2.6) -- (3.4,2.6);
\draw[black, line width=1] (3.4,1.9) -- (3.4,2.6);

\draw[black, line width=1] (4.4,1.9) -- (4.4,2.6);
\draw[black, line width=1] (4.4,1.9) -- (5.1,1.9);
\draw[black, line width=1] (4.4,2.6) -- (5.1,2.6);
\draw[black, line width=1] (5.1,1.9) -- (5.1,2.6);

\draw[black, line width=1] (3.4,2.25) -- (4.4,2.25);
\draw[red, photon, line width=1] (1.7,2.25) -- (2.7,2.25);
\draw[blue, gluon, line width=1] (1.7,0.55) -- (2.7,0.55);
\draw[black, line width=1] (1.35,1.9) -- (1.35,0.9);

\draw[black, line width=1] (5.7,1.75) -- (6.5,1.75);
\draw[red, photon, line width=1] (5.7,1.15) -- (6.5,1.15);
\draw[blue, gluon, line width=1] (5.7,0.55) -- (6.5,0.55);

\filldraw (6.9,2.35) node {$t$};
\filldraw (6.9,1.75) node {$W_1$};
\filldraw (6.9,1.15) node {$W_2$};
\filldraw (6.9,0.55) node {$W_3$};

\end{tikzpicture}

\end{center}
\vspace{-0.1in}
  \caption{\small An example graph $G \in \mathcal{G}_{C<4/3}$, where a normal $2$-color node $V_1$ is connected to a special $2$-color node $V_4$ and $V_1, V_4$ are connected to different types of edges.
  }
  \label{fig:necessary2}
\end{figure}
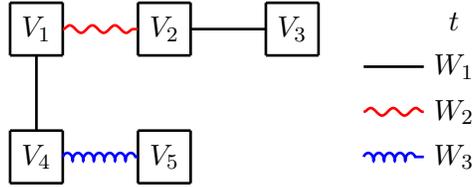

The set of graphs that satisfy the conditions in Theorem \ref{thm:necessary2} is denoted as $\mathcal{G}_{C < 3/4}^{\mbox{\tiny Thm \ref{thm:necessary2}}}$. The first two conditions are easily seen and an example for the third condition is shown in Fig.~\ref{fig:necessary2}. The proof of Theorem \ref{thm:necessary2} is deferred to Section \ref{sec:necessary2}. We give an intuitive explanation here on why $R < 4/3$ for the graph $G$ in Fig.~\ref{fig:necessary2}. From $(V_1, V_4, V_5)$, we can decode $W_1, W_3$, i.e., $2L_w$ bits. $V_4, V_5$ can contribute at most $2 L_v$ bits on $W_1, W_3$ so that the remaining $2L_w - 2L_v$ bits must come from $V_1$, leaving only $L_v - (2L_w - 2L_v) = 3L_v - 2L_w$ bits of room for $W_2$. Similarly, $V_2$ has at most $L_v - (L_w - L_v) = 2L_v - L_w$ bits of room for $W_2$ because at least $L_w - L_v$ bits of $W_1$ must come from $V_2$ (consider the $W_1$-edge $\{V_2, V_3\}$). The $W_2$-edge $\{V_1, V_2\}$ needs to have at least $L_w$ bits of room for the desired source $W_2$, i.e., $(3L_v - 2L_w) + (2L_v - L_w) \geq L_w$ so that $R = L_w/L_v \leq 5/4 < 4/3$.

\bigskip
Interestingly, if we exclude the graphs in $\mathcal{G}_{C < 3/4}^{\mbox{\tiny Thm \ref{thm:necessary2}}}$, i.e., those for which rate $4/3$ cannot be achieved, then the sufficient condition in Theorem \ref{thm:sufficient} generalizes immediately to more than $K=2$ source symbols. This result is stated in the following theorem.

\begin{theorem}\label{thm:sufficient2} [Sufficient Condition of $\mathcal{G}_{C=4/3}$]
A graph $G \in \mathcal{G}_{C \geq 4/3}$ if $G \notin \mathcal{G}_{C < 3/4}^{\mbox{\tiny Thm \ref{thm:necessary2}}}$ and $G$ contains no internal edge or for any internal edge, its residing path contains a $1$-color node.
\end{theorem}

The code construction of Theorem \ref{thm:sufficient2} is almost identical to that of Theorem \ref{thm:sufficient} and it turns out to work as long as the structures in Theorem \ref{thm:necessary2} are avoided. The detailed proof is deferred to Section \ref{sec:sufficient2} and an example is shown in Fig.~\ref{fig:sufficient2} to illustrate the idea. The assignment is still interference based, i.e., for each source symbol $W_k$, decompose all nodes connected to $W_k$-edges according to $W_{k'}$-components, where $k' \neq k$ (each node will belong only to one such component) and assign the same $W_k$ symbols within the same $W_{k'}$-components. After this operation, the interference dimension is controlled; the absence of internal edges (after removing $1$-color nodes) will guarantee the independence of desired symbols. The condition $G \notin \mathcal{G}_{C < 3/4}^{\mbox{\tiny Thm \ref{thm:necessary2}}}$ helps to guarantee that if a pair of connecting $2$-color nodes are connected to edges associated with more than $2$ source symbols (e.g., in Fig.~\ref{fig:sufficient2}, $\{V_2, V_3\}$ are associated with $3$ sources), the pair of nodes must be special $2$-color and the interference caused by the two interfering source symbols is still limited to $L_w/2$ dimensions (e.g., in Fig.~\ref{fig:sufficient2}, $V_2, V_3$ are special $2$-color and for $W_2$-edge $\{V_2,V_3\}$, the interference is one $A_i$ combination in $V_2$ and one $C_m$ combination in $V_3$, i.e., two dimensions in total). Other edges are the same as those in Theorem \ref{thm:sufficient} and decoding constraints (\ref{dec}) hold (see Fig.~\ref{fig:sufficient2}).

\vspace{0.1in}
\tikzset{
    photon/.style={decorate, decoration={snake,
      amplitude = 0.5mm,
      segment length = 3mm}, draw=red},
    electron/.style={draw=blue, postaction={decorate},
    decoration={markings,mark=at position .55 with {\arrow[draw=blue]{>}}}},
    gluon/.style={decorate, draw=blue,
        decoration={coil,amplitude=1.5pt, segment length=4pt}} 
}
\begin{figure}[H]
\begin{center}
\begin{tikzpicture}

\filldraw (1.35,3.25) node {$V_1$};
\filldraw (-0.2,3.25) node {$(A_1,A_2,A_3)$};

\filldraw (3.55,3.25) node {$V_2$};
\filldraw (2.6,4.35) node {$(B_9+B_{10},B_9+2B_{10},$};
\filldraw (3.5,3.95) node {$A_6+3A_7)$};

\filldraw (5.75,3.25) node {$V_3$};
\filldraw (7,4.35) node {$(B_{11}+B_{12},B_{11}+2B_{12},$};
\filldraw (5.9,3.95) node {$C_4+C_5)$};

\filldraw (7.95,3.25) node {$V_4$};
\filldraw (10.35,3.45) node {$(B_{13}+B_{14},C_4+3C_5,$};
\filldraw (10.7,3.05) node {$B_{13}+2B_{14}+C_4+4C_5)$};

\filldraw (1.35,1.05) node {$V_5$};
\filldraw (-0.25,1.45) node {$(A_4+A_5,$};
\filldraw (-0.08,1.05) node {$A_4+2A_5,$};
\filldraw (-0.15,0.65) node {$B_7+B_8)$};

\filldraw (3.55,1.05) node {$V_6$};
\filldraw (3.35,0.4) node {$(A_6+A_7,$};
\filldraw (3.52,0) node {$B_7+3B_8,$};
\filldraw (3.65,-0.4) node {$A_6+2A_7\,+$};
\filldraw (3.53,-0.8) node {$B_7+4B_8)$};

\filldraw (5.75,1.05) node {$V_7$};
\filldraw (5.75,0.4) node {$(C_1,C_2,C_3)$};

\filldraw (7.95,1.05) node {$V_8$};
\filldraw (9.3,1.45) node {$(C_6+C_7,$};
\filldraw (9.47,1.05) node {$C_6+2C_7,$};
\filldraw (9.5,0.65) node {$B_{13}+3B_{14})$};

\filldraw (1.35,-1.15) node {$V_9$};
\filldraw (-0.2,-1.15) node {$(B_1,B_2,B_3)$};

\filldraw (5.75,-1.15) node {$V_{10}$};
\filldraw (7.4,-1.15) node {$(B_4,B_5,B_6)$};

\draw[black, line width=1] (1,-1.5) -- (1,-0.8);
\draw[black, line width=1] (1,-1.5) -- (1.7,-1.5);
\draw[black, line width=1] (1,-0.8) -- (1.7,-0.8);
\draw[black, line width=1] (1.7,-1.5) -- (1.7,-0.8);

\draw[black, line width=1] (1,0.7) -- (1,1.4);
\draw[black, line width=1] (1,0.7) -- (1.7,0.7);
\draw[black, line width=1] (1,1.4) -- (1.7,1.4);
\draw[black, line width=1] (1.7,0.7) -- (1.7,1.4);

\draw[black, line width=1] (3.2,0.7) -- (3.2,1.4);
\draw[black, line width=1] (3.2,0.7) -- (3.9,0.7);
\draw[black, line width=1] (3.2,1.4) -- (3.9,1.4);
\draw[black, line width=1] (3.9,0.7) -- (3.9,1.4);

\draw[black, line width=1] (1,2.9) -- (1,3.6);
\draw[black, line width=1] (1,2.9) -- (1.7,2.9);
\draw[black, line width=1] (1,3.6) -- (1.7,3.6);
\draw[black, line width=1] (1.7,2.9) -- (1.7,3.6);

\draw[black, line width=1] (3.2,2.9) -- (3.2,3.6);
\draw[black, line width=1] (3.2,2.9) -- (3.9,2.9);
\draw[black, line width=1] (3.2,3.6) -- (3.9,3.6);
\draw[black, line width=1] (3.9,2.9) -- (3.9,3.6);

\draw[black, line width=1] (5.4,-1.5) -- (5.4,-0.8);
\draw[black, line width=1] (5.4,-1.5) -- (6.1,-1.5);
\draw[black, line width=1] (5.4,-0.8) -- (6.1,-0.8);
\draw[black, line width=1] (6.1,-1.5) -- (6.1,-0.8);

\draw[black, line width=1] (5.4,0.7) -- (5.4,1.4);
\draw[black, line width=1] (5.4,0.7) -- (6.1,0.7);
\draw[black, line width=1] (5.4,1.4) -- (6.1,1.4);
\draw[black, line width=1] (6.1,0.7) -- (6.1,1.4);

\draw[black, line width=1] (7.6,0.7) -- (7.6,1.4);
\draw[black, line width=1] (7.6,0.7) -- (8.3,0.7);
\draw[black, line width=1] (7.6,1.4) -- (8.3,1.4);
\draw[black, line width=1] (8.3,0.7) -- (8.3,1.4);

\draw[black, line width=1] (5.4,2.9) -- (5.4,3.6);
\draw[black, line width=1] (5.4,2.9) -- (6.1,2.9);
\draw[black, line width=1] (5.4,3.6) -- (6.1,3.6);
\draw[black, line width=1] (6.1,2.9) -- (6.1,3.6);

\draw[black, line width=1] (7.6,2.9) -- (7.6,3.6);
\draw[black, line width=1] (7.6,2.9) -- (8.3,2.9);
\draw[black, line width=1] (7.6,3.6) -- (8.3,3.6);
\draw[black, line width=1] (8.3,2.9) -- (8.3,3.6);

\draw[black, line width=1] (1.7,3.25) -- (3.2,3.25);
\draw[black, line width=1] (1.35,2.9) -- (1.35,1.4);
\draw[black, line width=1] (1.7,1.05) -- (3.2,1.05);

\draw[red, photon, line width=1] (1.35,0.7) -- (1.35,-0.8);
\draw[red, photon, line width=1] (3.55,2.9) -- (3.55,1.4);
\draw[red, photon, line width=1] (3.9,3.25) -- (5.4,3.25);
\draw[red, photon, line width=1] (6.1,3.25) -- (7.6,3.25);

\draw[red, photon, line width=1] (3.9,1.05) -- (5.4,-1.15);
\draw[red, photon, line width=1] (7.95,0.7) -- (6.1,-1.15);

\draw[blue, gluon, line width=1] (5.75,2.9) -- (5.75,1.4);
\draw[blue, gluon, line width=1] (7.95,2.9) -- (7.95,1.4);
\draw[blue, gluon, line width=1] (7.95,2.9) -- (6.1,1.05);

\draw[black, line width=1] (11.1,1) -- (11.9,1);
 \draw[red, photon, line width=1] (11.1,0) -- (11.9,0);
 \draw[blue, gluon, line width=1] (11.1,-1) -- (11.9,-1);

\filldraw (12.3,1.8) node {$t$};
\filldraw (12.3,1) node {$W_1$};
\filldraw (12.3,0) node {$W_2$};
\filldraw (12.3,-1) node {$W_3$};

\end{tikzpicture}

\end{center}
\vspace{-0.1in}
  \caption{\small An example graph $G \in \mathcal{G}_{C \geq 4/3}$ with $K=3$ source symbols and code construction for rate $4/3$. $W_1 = (a_1, \cdots,a_4), W_2 = (b_1, \cdots,b_4), W_3 = (c_1, \cdots, c_4)$ and each $A_i\backslash B_j \backslash C_m$ is a generic linear combination of $(a_1,\cdots,a_4)\backslash (b_1,\cdots,b_4) \backslash (c_1, \cdots, c_4)$.}
\label{fig:sufficient2}
\end{figure}
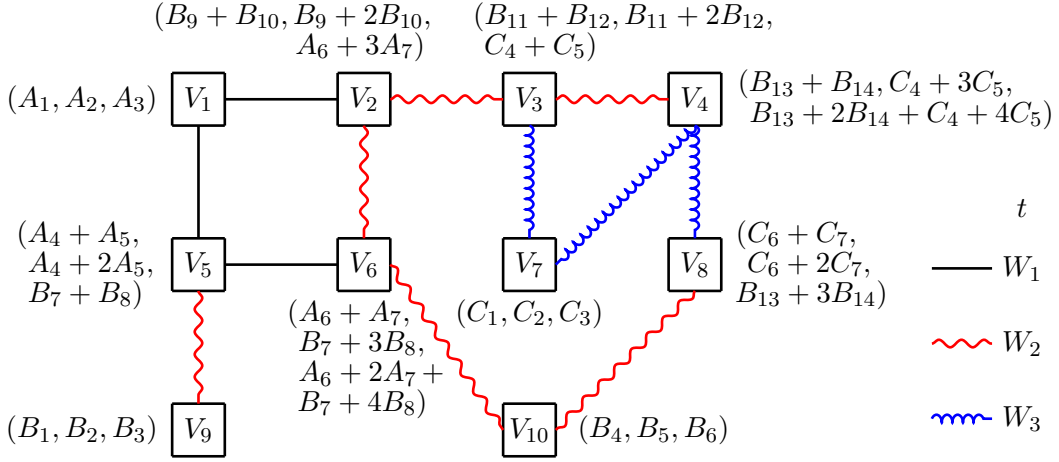

\section{Proofs}
\subsection{Proof of Theorem \ref{thm:C2}: $\mathcal{G}_{C=2}, \mathcal{G}_{C=3/2}$} \label{sec:C2}
In this section, we provide the full characterization of $\mathcal{G}_{C=2}$ and $\mathcal{G}_{C=3/2}$. From the proof, we can obtain that the three highest 
storage code capacity values are $2, 3/2, 4/3$.

\subsubsection{If and only if condition of $
\mathcal{G}_{C=2}$}
We show that $G(\mathcal{V},\mathcal{E},t) \in \mathcal{G}_{C=2}$ if and only if every node $V \in \mathcal{V}$ is $1$-color. We prove the if part and the only if part sequentially.

{\bf If part:} If every node $V \in \mathcal{V}$ is $1$-color, then we prove that the capacity is $2$. First, we show that $R \leq 2$. Note that we assume $G$ has no isolated nodes, so $G$ must contain one edge, say $W_k$-edge $\{V_i,V_j\}$. From the decoding constraint (\ref{dec}), we have
\begin{eqnarray}
&&L_w \overset{(\ref{h1})}{=} H(W_k) \overset{(\ref{dec})}{=} I(V_i,V_j;W_k) \leq H(V_i,V_j) \leq 2L_v \label{eq:c21}\\
\Rightarrow && R \overset{(\ref{rate})}{=} L_w/L_v \leq 2
\end{eqnarray}
where the last inequality in (\ref{eq:c21}) follows from the fact that each coded symbol $V_i$ contains at most $L_v$ bits.

Second, we show that symbol rate $R = 2$ is achievable, by an MDS code. Note that each node is $1$-color; suppose there are $M_k$ nodes that are only connected to $W_k$-edges, $k\in[K]$ and denote this set of nodes by $\mathcal{V}_k$. Choose the field size $p$ to be a prime that is no smaller than $\max_{k\in[K]} M_k$. Set $L_w = 2\log_2 p, L_v = \log_2 p$ so that each source$\backslash$coded symbol is comprised of $2\backslash1$ symbols$\backslash$symbol from $\mathbb{F}_p$ and the rate achieved is $2$. Generate MDS coded symbols as follows.
\begin{eqnarray}
W_k &=& (W_k(1); W_k(2)) \in \mathbb{F}_{p}^{2 \times 1} \\
X_k &=& (X_k(1); \cdots; X_k(M_k)) \triangleq {\bf V}_k W_k  \in \mathbb{F}_p^{M_k \times 1}
\end{eqnarray}
where ${\bf V}_k \in \mathbb{F}_p^{M_k \times 2}$ is a full rank Vandermonde matrix so that from any two elements of $X_k$, we can recover $W_k$ (i.e., MDS). Finally, we assign each node in $\mathcal{V}_k$ a distinct element of $X_k$ so that from any $W_k$-edge, we can decode $W_k$.

{\bf Only if part:}
We show that if there exists an $M$-color node, $M \geq 2$, then $R < 2$ so that the capacity must be strictly smaller than $2$ and further, the capacity will drop to $3/2$ at least. Suppose we have an $M$-color node $V_i$ that is connected to $W_{k_1}$-edge $\{V_i,V_{j_1}\},\cdots,W_{k_M}$-edge $\{V_i,V_{j_M}\}$, then
\begin{eqnarray}
M L_w &\overset{(\ref{h1})}{=}& H(W_1,\cdots,W_k) \\
&\overset{(\ref{dec})}{=}& I(V_i,V_{j_1},\cdots,V_{j_M};W_1,\cdots,W_k) \\
&\leq& H(V_i,V_{j_1},\cdots,V_{j_M}) \\
&\leq& (M+1) L_v \label{kcolor}\\
\Rightarrow~ R &\overset{(\ref{rate})}{=}& L_w/L_v \leq (M+1)/M \leq 3/2 < 2.
\label{eq:2/3}
\end{eqnarray}

\subsubsection{If and only if condition of $
\mathcal{G}_{C=3/2}$}
We show that $G(\mathcal{V},\mathcal{E},t) \in \mathcal{G}_{C=3/2}$ if and only if every node $V \in \mathcal{V}$ is $2$-color or $1$-color (and there exists a $2$-color node) and there are no connected $2$-color nodes. We prove the if part and the only if part sequentially.

{\bf If part:} We show that $C = 3/2$ if the condition above is satisfied.
First, $G$ contains a $2$-color node, so from (\ref{eq:2/3}) we have that $R \leq 3/2$. 
Second, we show that $R=3/2$ is achievable, again by an MDS code. 
Note that each node is $1$-color or $2$-color. 
Consider the nodes that are connected to $W_k$-edges, among which suppose $M^1_k$ are $1$-color (denote this set by $\mathcal{V}^1_k$) and $M_k^2$ are $2$-color (denote this set by $\mathcal{V}^2_k$).
Choose the field size $p$ to be a prime that is no smaller than $\max_{k} (2M^1_k+M^2_k$). Set $L_w = 3\log_2 p, L_v = 2\log_2 p$, i.e., each source$\backslash$coded symbol is comprised of $3\backslash 2$ symbols from $\mathbb{F}_p$ and the rate achieved is $3/2$. Generate MDS coded symbols as follows.
\begin{eqnarray}
W_k &=& (W_k(1); W_k(2); W_k(3)) \in \mathbb{F}_{p}^{3 \times 1} \\
X_k &=& (X_k(1); \cdots; X_k(2M^1_k + M^2_k)) \triangleq {\bf V}_k W_k  \in \mathbb{F}_p^{(2M^1_k + M^2_k) \times 1}
\end{eqnarray}
where ${\bf V}_k \in \mathbb{F}_p^{(2M^1_k + M^2_k) \times 3}$ is a full rank Vandermonde matrix so that from any three elements of $X_k$, we can recover $W_k$ (i.e., MDS). The existence of such a full rank Vandermonde matrix is guaranteed due to our field size choice. Finally, we assign each node in $\mathcal{V}_k^1 \backslash \mathcal{V}_k^2$ two$\backslash$one distinct elements$\backslash$element of $X_k$. Note that any node $V \in \mathcal{V}$ will be assigned two $\mathbb{F}_p$ symbols. To verify that the decoding constraint (\ref{dec}) holds, consider any $W_k$-edge $\{V_i, V_j\}$, where $V_i, V_j$ cannot both be $2$-color because from our condition of $\mathcal{G}_{C=3/2}$, $2$-color nodes do not connect. As a $2$-color node contains one element of $X_k$ and a $1$-color node contains two elements of $X_k$, $(V_i, V_j)$ will contain at least three elements of $X_k$, from which we can recover $W_k$.

{\bf Only if part:}
 We show that $C \neq 3/2$ if the condition of $\mathcal{G}_{C = 3/2}$ is violated, i.e., if 1) there only exist $1$-color nodes, 2) there is an $M$-color node, where $M \geq 3$, or 3) if there are connected $2$-color nodes, say $V_i, V_j$. For Case 1), $G \in \mathcal{G}_{C=1}$; for Case 2), from (\ref{eq:2/3}) we have $R \leq 4/3 < 3/2$; for Case 3), we next show that $R \leq 4/3$ so that the proof is complete and when capacity $2$ and $3/2$ cannot be achieved, it drops to $4/3$ at least (and rate $4/3$ is achievable for some graph, e.g., Theorem \ref{thm:sufficient}, so the third-highest capacity value is $4/3$).
 
Suppose $\{V_i, V_j\}$ is a $W_k$-edge. As $V_i$ is $2$-color, $V_i$ must be connected to some node $V_{i_1}, i_1 \neq i$ with a $W_{k_1}$-edge, where $k_1 \neq k$. Consider the $W_{k_1}$-edge $\{V_{i},V_{i_1}\}$ and we have
\begin{eqnarray}
L_w &\overset{(\ref{h1})}{=}& H(W_{k_1}) \label{eq:v1}\\
&\overset{(\ref{dec})}{=}& I(V_{i},V_{i_1};W_{k_1}) \\
&=& H(V_{i},V_{i_1}) - H(V_{i},V_{i_1} | W_{k_1}) \\
&\leq& 2L_v- H(V_{i},V_{i_1} | W_{k_1}) \\
\Rightarrow ~~ H(V_{i}|W_{k_1}) &\leq& H(V_{i},V_{i_1} | W_{k_1}) \leq 2L_v - L_w.\label{eq:v1lw}
\end{eqnarray}

Symmetrically, $V_j$ is $2$-color so that $V_j$ must be connected to $V_{j_1}, j_1 \neq j$ with a $W_{k_2}$-edge, where $k_2 \neq k$. Note that $j_1$ may be the same as $i_1$ and $k_2$ may be the same as $k_1$. The same proof will work under all circumstances. Consider the $W_{k_2}$-edge $\{V_{j},V_{j_1}\}$. Following the derivation of (\ref{eq:v1lw}), we have
\begin{eqnarray}
H(V_{j}|W_{k_2}) \leq 2L_v - L_w.\label{eq:v2lw}
\end{eqnarray}

Finally consider the $W_{k}$-edge $\{V_{i},V_{j}\}$ and we have
\begin{eqnarray}
L_w &\overset{(\ref{h1})}{=}& 
H(W_{k}|W_{k_1},W_{k_2}) \\
&\leq& H(W_{k},V_{i},V_{j}|W_{k_1},W_{k_2}) \\
&\overset{(\ref{dec})}{=}& H(V_{i},V_{j}|W_{k_1},W_{k_2}) \\
&\leq& H(V_{i}|W_{k_1})+ H(V_{j}|W_{k_2})\\
&\overset{(\ref{eq:v1lw})(\ref{eq:v2lw})}\leq& 2L_v - L_w + 2L_v - L_w \\
\Rightarrow R &\overset{(\ref{rate})}{=}& L_w/L_v \leq 4/3.
\end{eqnarray}

\subsection{Proof of Theorem \ref{thm:sufficient}: Sufficient Condition of $\mathcal{G}_{C=4/3}$ with $K=2$} \label{sec:sufficient}
We show that if a graph\footnote{For simplicity, the edge association mapping $t$ is omitted from the graph notation $G$ in this section.} $G(\mathcal{V},\mathcal{E})$ contains no internal edge or each residing path contains one $1$-color node, then $R = 4/3$ is achievable. We first present the code construction and then prove it satisfies the decoding constraint (\ref{dec}).

\subsubsection{Code construction}\label{sec:code}
Choose the field size $p$ to be a prime that is greater than $4|\mathcal{E}|$. 
Set $L_w = 4\log_2 p, L_v = 3 \log_2 p$ so that each source$\backslash$coded symbol is comprised of $4\backslash 3$ symbols from $\mathbb{F}_p$ and the rate achieved is $4/3$. 

Consider the set of nodes that are connected to $W_k$-edges, $k \in \{1,2\}$ and denote this set by $\mathcal{V}_k$. 
Consider the subgraph of $G(\mathcal{V}, \mathcal{E})$ whose node set is $\mathcal{V}_k$ and edge set is comprised of all $W_{3-k}$-edges that are connected to some node in $\mathcal{V}_k$, denoted by $\mathcal{E}_{3-k}$ and denote this subgraph by ${G}_k(\mathcal{V}_k, \mathcal{E}_{3-k})$. Decompose ${G}_k(\mathcal{V}_k, \mathcal{E}_{3-k})$ into $W_{3-k}$-components and suppose we have $M_k$ such components. Among these $M_k$ $W_{3-k}$-components, suppose $M_k^1$ components are comprised of $1$-color nodes (each such component is an isolated node) and label them as $P_k^{[1]}, \cdots, P_k^{[M_k^1]}$; the remaining $M_k^2 = M_k - M_k^1$ components are comprised of $2$-color nodes and label them as $Q_k^{[1]}, \cdots, Q_k^{[M_k^2]}$. For an example of subgraph ${G}_k(\mathcal{V}_k, \mathcal{E}_{3-k})$ and its decomposition, refer to Fig.~\ref{fig:sufficient_component}.

Generate generic linear combinations of the source symbols as follows.
\begin{eqnarray}
W_k &=& (W_k(1); \cdots; W_k(4)) \in \mathbb{F}_{p}^{4 \times 1}, k \in \{1,2\} \\
X_k &=& (X_k(1); \cdots; X_k(3M_k^1 + 2M_k^2)) \triangleq {\bf H}_k W_k  \in \mathbb{F}_p^{(3M_k^1 + 2M_k^2) \times 1} \label{eq:xk}
\end{eqnarray}
where ${\bf H}_k$ is a $(3M_k^1 + 2M_k^2) \times 4$ matrix over the field $\mathbb{F}_p$ and each element of ${\bf H}_k$ is chosen uniformly and independently from $\mathbb{F}_p$. Thus our construction is randomized and we will show that the probability that all decoding constraints (\ref{dec}) are satisfied is strictly larger than $0$ so that one feasible code construction exists.

Consider $P_k^{[m]}, m \in [M_k^1]$, denote its node by $V$, and set
\begin{eqnarray}
V = \big(X_k(3m-2), X_k(3m-1), X_k(3m) \big). \label{eq:1}
\end{eqnarray}
This step completes the assignment for all $1$-color nodes, each of which must reside in one $P_k^{[m]}$.

Consider $Q_k^{[m]}, m \in [M_k^2]$. Suppose $Q_k^{[m]}$ contains $J$ nodes $V_{i_1}, \cdots, V_{i_J}$ (each must be $2$-color). Consider node $V_{i_j}, j \in [J]$.
\begin{eqnarray}
&& \mbox{If $V_{i_j}$ is $W_k$-special, set $V_{i_j}^{[k]} \triangleq X_k\big(3M_k^1 + 2m-1 \big) + (2j-1) X_k\big(3M_k^1 + 2m \big) $;} \\
&& \mbox{otherwise, set $V_{i_j}^{[k]} = (V_{i_j}^{[k]}(1), V_{i_j}^{[k]}(2)) \triangleq \Big( X_k\big(3M_k^1 + 2m-1 \big) + (2j-1) X_k\big(3M_k^1 + 2m \big),$} \notag\\
&& ~~~~~~~~~~~~~~~~~~~~~~~~~~~~~~~~~~~~~~~~~~~~~~~~~~~~~~~\mbox{$X_k\big(3M_k^1 + 2m-1 \big) + 2j X_k\big(3M_k^1 + 2m \big) \Big) $.} \label{eq:2s}
\end{eqnarray}

Finally, after setting $V^{[1]}, V^{[2]}$ for each $2$-color node $V$, we are ready to set $V$.
\begin{eqnarray}
&& \mbox{If $V$ is normal, then set $V = (V^{[1]}(1), V^{[2]}(1), V^{[1]}(2) + V^{[2]}(2))$;} \\
&& \mbox{otherwise, set $V = (V^{[1]}, V^{[2]})$.} \label{eq:2}
\end{eqnarray}
Note that for a special $2$-color node, at least one of $V^{[1]}, V^{[2]}$ will be one symbol so that $V$ will contain no more than three symbols (when a node $V$ is simultaneously $W_1$-special and $W_2$-special, $V$ will have only two symbols and we may zero-pad to make its length three). This completes the assignment for all $2$-color nodes and the code construction is complete.

\subsubsection{Proof of correctness}\label{sec:corr}
We show that the decoding constraint (\ref{dec}) is satisfied. Consider any edge $\{V_i, V_j\} \in \mathcal{E}$ and suppose it is a $W_k$-edge.

When $V_i, V_j$ contain one $1$-color code, say $V_i$, then $V_i$ contains three elements of $X_k$ (say, $X_k(m_1), X_k(m_1+1), X_k(m_1+2)$; refer to (\ref{eq:1})) and $V_j$ contains at least one generic combination of distinct two elements of $X_k$ or one distinct element of $X_k$ (say, $X_k(m_2) + j X_k(m_2+1)$ or $X_k(m_2)$; refer to (\ref{eq:1}) - (\ref{eq:2})). These $4$ symbols in $X_k$ can be written as a multiplication of a $4 \times 4$ matrix, denoted by ${\bf T}_{ij}$ and the source symbol vector $W_k$. View the determinant of ${\bf T}_{ij}$ as a polynomial $T_{ij}({\bf H}_1, {\bf H}_2)$, whose variables are the elements of ${\bf H}_1, {\bf H}_2$ (refer to (\ref{eq:xk})). $T_{ij}({\bf H}_1, {\bf H}_2)$ is not a zero-polynomial as we may set $X_k(m_1) = W_k(1), X_k(m_1+1) = W_k(2), X_k(m_1+2) = W_k(3), X_k(m_2) = W_k(4), X_k(m_2+1) = 0$ so that ${\bf T}_{ij}$ is an identity matrix and $T_{ij}({\bf H}_1, {\bf H}_2)$ = 1.

We are left with cases where $V_i, V_j$ are both $2$-color. Note that $V_i, V_j$ cannot be $W_k$-special. We have three cases.
\begin{enumerate}
    \item $V_i, V_j$ are both $W_{3-k}$-special. Then each of $V_i, V_j$ contains two generic combinations of two distinct elements of $X_k$ (refer to (\ref{eq:2s}) and (\ref{eq:2})), say
    \begin{eqnarray}
    && X_k(m_1) + (2j_1-1) X_k(m_1 + 1), X_k(m_1) + 2j_1 X_k(m_1 + 1) ~\mbox{from}~V_i~\mbox{and} \notag\\
    && X_k(m_2) + (2j_2-1) X_k(m_2 + 1), X_k(m_2) + 2j_2 X_k(m_2 + 1) ~\mbox{from}~V_j \label{eq:2ss}
    \end{eqnarray}
    where the elements of $X_k$ are all distinct because $V_i, V_j$ belong to different $W_{3-k}$-components in the decomposition of ${G}_k(\mathcal{V}_k, \mathcal{E}_{3-k})$. Otherwise, $\{V_i, V_j\}$ is an internal edge after removing $1$-color nodes, which contradicts the condition of Theorem \ref{thm:sufficient}. From the four symbols in (\ref{eq:2ss}), we can recover four distinct elements of $X_k$, i.e., $(X_k(m_1); X_k(m_1+1); X_k(m_2); X_k(m_2+1))$, which can be similarly written as ${\bf T}_{ij}^{4 \times 4} W_k$. View $\det({\bf T}_{ij})$ as a polynomial $T_{ij}({\bf H}_1, {\bf H}_2)$, which is not the zero-polynomial.
    
    \item One of $V_i, V_j$ is $W_{3-k}$-special, say $V_i$ and the other is normal, say $V_j$. From (\ref{eq:2s}) - (\ref{eq:2}), we have
    \begin{eqnarray}
     V_i &=& \big( X_k(m_1) + (2j_1-1) X_k(m_1 + 1), X_k(m_1) + 2j_1 X_k(m_1 + 1), \notag\\
     && ~X_{3-k}(m^*) + (2i_1-1) X_{3-k}(m^*+1) \big) \notag \\
     V_j &=& \big( X_k(m_2) + (2j_2-1) X_k(m_2 + 1), X_{3-k}(m^*) + (2i_2-1) X_{3-k}(m^* + 1),  \notag \\
     && ~X_k(m_2) + 2j_2 X_k(m_2 + 1) + X_{3-k}(m^*) + 2i_2 X_{3-k}(m^* + 1) \big) \label{eq:21}
    \end{eqnarray}
    where the elements of $X_k$ are all distinct due to the same reason as above; the elements of $X_{3-k}$ must be the same, i.e., $m^*$ appears in both $V_i, V_j$ because $\{V_i, V_j\}$ is a $W_k$-edge so that $V_i, V_j$ belong to the same $W_k$-component in the decomposition of ${G}_{3-k}(\mathcal{V}_{3-k}, \mathcal{E}_{k})$. Further, $V_i, V_j$ are distinct so $i_1 \neq i_2$ in (\ref{eq:21}) (refer to (\ref{eq:2s})). Thus from $(V_i, V_j)$, we can first decode and remove $X_{3-k}(m^*), X_{3-k}(m^*+1)$, leaving us with four distinct elements of $X_k$, i.e., $(X_k(m_1); X_k(m_1+1); X_k(m_2); X_k(m_2+1)) = {\bf T}_{ij}^{4 \times 4} W_k$. View $\det({\bf T}_{ij})$ as a polynomial $T_{ij}({\bf H}_1, {\bf H}_2)$, which is non-zero.
    
    \item $V_i, V_j$ are both normal. From (\ref{eq:2s}) - (\ref{eq:2}), we have
    \begin{eqnarray}
     V_i &=& \big( X_k(m_1) + (2j_1-1) X_k(m_1 + 1), X_{3-k}(m^*) + (2i_1-1) X_{3-k}(m^* + 1),  \notag \\
     && ~X_k(m_1) + 2j_1 X_k(m_1 + 1) + X_{3-k}(m^*) + 2i_1 X_{3-k}(m^* + 1) \notag \\
     V_j &=& \big( X_k(m_2) + (2j_2-1) X_k(m_2 + 1), X_{3-k}(m^*) + (2i_2-1) X_{3-k}(m^* + 1),  \notag \\
     && ~X_k(m_2) + 2j_2 X_k(m_2 + 1) + X_{3-k}(m^*) + 2i_2 X_{3-k}(m^* + 1) \big)
    \end{eqnarray}
    where the elements of $X_k$ are all distinct, the elements of $X_{3-k}$ must be the same, and $i_1 \neq i_2$. Thus from $(V_i, V_j)$, we can first decode and remove $X_{3-k}(m^*), X_{3-k}(m^*+1)$, leaving us with $(X_k(m_1); X_k(m_1+1); X_k(m_2); X_k(m_2+1)) = {\bf T}_{ij}^{4 \times 4} W_k$ and $\det({\bf T}_{ij})$ is a non-zero polynomial $T_{ij}({\bf H}_1, {\bf H}_2)$.
\end{enumerate}

Finally, consider all edges of $G(\mathcal{V}, \mathcal{E})$ and consider $\prod_{i,j: \{V_i, V_j\} \in \mathcal{E}} T_{ij}({\bf H}_1, {\bf H}_2)$, which is a polynomial with degree at most $4 |\mathcal{E}|$. Now each element of ${\bf H}_1, {\bf H}_2$ is selected independently and uniformly from $\mathbb{F}_p$, where $p > 4 |\mathcal{E}|$. By the Schwartz–Zippel lemma \cite{Demillo_Lipton, Schwartz, Zippel}, we have
\begin{eqnarray}
\Pr\left(\prod_{i,j:\{V_i, V_j\} \in \mathcal{E}} T_{ij}({\bf H}_1, {\bf H}_2) = 0 \right) \leq \frac{4|\mathcal{E}|}{p} < 1. \label{eq:sz}
\end{eqnarray}
Therefore there exists a realization of ${\bf H}_1, {\bf H}_2$ so that each $T_{ij}({\bf H}_1, {\bf H}_2) \neq 0$ and each ${\bf T}_{ij}$ has full rank, i.e., $W_k$ can be recovered from $\{V_i, V_j\}$ and all decoding constraints (\ref{dec}) are satisfied.

\subsection{Proof of Theorem \ref{thm:necessary}: Necessary Condition of $\mathcal{G}_{C=4/3}$ with $K=2$} \label{sec:necessary}
We show that $R = 4/3$ cannot be achieved if a graph $G$ contains an internal $W_k$-edge $\{V_{i_1}, V_{i_P}\}$ and its residing $W_{3-k}$-path $(\{V_{i_1}, V_{i_2}\}, \cdots, \{V_{i_{P-1}}, V_{i_P}\})$ contains only $2$-color nodes, $V_{i_1}, \cdots, V_{i_P}$, among which at most one is special (if there exists, suppose it is $V_{i_p}, 1 < p < P$). 

To set up the proof by contradiction, let us assume that $R=\lim_{L_w \rightarrow \infty} L_w/L_v=4/3$ is asymptotically achievable (the same proof works for the exact achievable case by replacing $o(L_w)$ with zero), i.e., 
\begin{eqnarray}
L_v=(3L_w)/4 + o(L_w). \label{eq:r4/3}
\end{eqnarray}
We show that a $2$-color node $V$ must contain at least $L_w/4$ bits (minimum amount of desired information) and at most $L_w/2$ bits (maximum amount of interference) of information about each of $W_1$ and $W_2$. This result is stated in the following lemma.

\begin{lemma}[$2$-color Node] \label{lemma:2color}
When $R=4/3$, for any $2$-color node $V$, we have
\begin{eqnarray}
L_w/4 + o(L_w) ~\leq~ H(V|W_k) &\leq& L_w/2 + o(L_w), ~\forall k \in \{1,2\}. \label{eq:2color}
\end{eqnarray}
\end{lemma}

{\it Proof:} As $V$ is $2$-color, we have a $W_1$-edge $\{V, V_{j_1}\}$ and a $W_2$-edge $\{V, V_{j_2}\}$. 
We prove (\ref{eq:2color}) when $k=1$ and the proof when $k=2$ follows from symmetry.

Consider the $W_1$-edge $\{V,V_{j_1}\}$. Following the steps in (\ref{eq:v1}) to (\ref{eq:v1lw}), we have
\begin{eqnarray}
H(V |W_1) \leq 2L_v - L_w \overset{(\ref{eq:r4/3})}{=} L_w/2 + o(L_w).
\end{eqnarray}

Consider the $W_2$-edge $\{V,V_{j_2}\}$. We have
\begin{eqnarray}
L_w &\overset{(\ref{h1})}{=}& H(W_2|W_1) \\
&\leq& H(W_2,V_{},V_{j_2}|W_1) \\
&\overset{(\ref{dec})}{=}& H(V_{},V_{j_2}|W_1) \\
&\leq& H(V |W_1)+H(V_{j_2}) \\
&\leq& H(V |W_1)+ L_v \\
\Rightarrow~~ H(V |W_1) &\geq& L_w - L_v \overset{(\ref{eq:r4/3})}{=} L_w/4 + o(L_w).
\end{eqnarray}

\hfill\QED

Next, we tighten the result in Lemma \ref{lemma:2color} when the $2$-color node is further normal. Specifically, a normal $2$-color node $V$ must contain exactly $L_w/2$ bits of information about each of $W_1$ and $W_2$. This result is stated in the following lemma.
\begin{lemma}[Normal $2$-color Node] \label{lemma:normal}
When $R=4/3$, for any normal $2$-color node $V$, we have
\begin{eqnarray}
H(V |W_k) = L_w/2 + o(L_w), ~\forall k \in \{1,2\}. \label{eq:normal}
\end{eqnarray}
\end{lemma}

{\it Proof:} As $V$ is normal $2$-color, it must be connected to $V_{j_1}$ through a $W_1$-edge and $V_{j_2}$ through a $W_2$-edge, and further $V_{j_1}, V_{j_2}$ are $2$-color. 
The `$\leq$' direction of (\ref{eq:normal}) has been proved in (\ref{eq:2color}), so we only need to prove the `$\geq$' direction, which is considered in the following when $k=2$ and the proof when $k=1$ follows from symmetry.

Consider the $W_1$-edge $\{V,V_{j_1}\}$. We have
\begin{eqnarray}
L_w 
&\overset{(\ref{h1})}{=}& H(W_1|W_2) \\
&\leq& H(W_1,V, V_{j_1}|W_2) \\
&\overset{(\ref{dec})}{=}& H(V, V_{j_1}|W_2) \\
&\leq& H(V|W_2)+H(V_{j_1}|W_2) \\
&\overset{(\ref{eq:2color})}{\leq}& H(V|W_2)+L_w/2 + o(L_w) \label{lem23}\\
\Rightarrow ~~ H(V|W_2) &\geq& L_w/2 + o(L_w)
\end{eqnarray}
where (\ref{lem23}) holds because $V_{j_1}$ is a $2$-color node so that we may apply (\ref{eq:2color}) of Lemma \ref{lemma:2color}.

\hfill\QED

After establishing the properties on the nodes, we proceed to consider the edges. We show that for any two connected $2$-color nodes, the interference contained in them is $L_w/2$ bits if the two nodes contain one normal $2$-color node.

\begin{lemma}[$W_{3-k}$-edge] \label{lemma:edge}
When $R=4/3$, for any $W_{3-k}$-edge $\{V_i,V_j\}$ where $V_i, V_j$ are $2$-color and at least one of $V_i, V_j$ is normal, we have
\begin{eqnarray}
H(V_i,V_j|W_{3-k}) = L_w/2 + o(L_w). \label{eq:edge} 
\end{eqnarray}
\end{lemma}

{\it Proof:} 
Suppose $V_i$ is normal, then on the one hand, we have
\begin{eqnarray}
H(V_i,V_j|W_{3-k}) \geq H(V_i|W_{3-k}) \overset{(\ref{eq:normal})}{=} L_w/2 + o(L_w). \label{lm21}
\end{eqnarray}

On the other hand, we have 
\begin{eqnarray}
H(V_i,V_j|W_{3-k}) \overset{(\ref{eq:v1lw})}{\leq} 2L_v-L_w \overset{(\ref{eq:r4/3})}{=} L_w/2 + o(L_w)
\end{eqnarray}
so that the proof is complete.

\hfill\QED

We now go from the properties of edges in Lemma \ref{lemma:edge} to those of paths that were made up of such edges. We show that the interference contained in a sequence of such edges, i.e., a path, is $L_w/2$ bits, in the following lemma.
\begin{lemma}[$W_{3-k}$-path] \label{lemma:path}
When $R=4/3$, for a $W_{3-k}$-path $(\{V_{i_1}, V_{i_2}\}, \cdots, \{V_{i_{P-1}}, V_{i_P} \} )$ where $V_{i_1}, \cdots, V_{i_{p-1}}, V_{i_{p+1}}, \cdots, V_{i_P}, 1 < p < P$ are normal and $V_{i_p}$ is either normal or special, we have
\begin{eqnarray}
H(V_{i_1},\cdots,V_{i_p}|W_{3-k}) &\leq& L_w/2 + o(L_w) \label{eq:path}\\
H(V_{i_p},\cdots,V_{i_P}|W_{3-k}) &\leq& L_w/2 + o(L_w).\label{eq:path2}
\end{eqnarray}
\end{lemma}

{\it Proof:} We prove (\ref{eq:path}) and (\ref{eq:path2}) follows similarly. The proof is based on a straightforward application of the submodular property on the entropy function to (\ref{eq:edge}) in Lemma \ref{lemma:edge} (note that each edge in the path contains at most one special $2$-color node).
\begin{eqnarray}
&& H(V_{i_1},V_{i_2}|W_{3-k})+\cdots+H(V_{i_{p-1}},V_{i_p}|W_{3-k}) \notag\\
&\geq& H(V_{i_1},\cdots,V_{i_p}|W_{3-k}) + H(V_{i_2}|W_{3-k})+\cdots+H(V_{i_{p-1}}|W_{3-k})  \\
\overset{(\ref{eq:normal})(\ref{eq:edge})}{\Longrightarrow} && (p-1) L_w/2 ~\geq~ H(V_{i_1},\cdots,V_{i_p}|W_{3-k}) + (p-2)L_w/2 + o(L_w) \\
\Longrightarrow~  && H(V_{i_1},\cdots,V_{i_p}|W_{3-k}) ~\leq~ L_w/2 + o(L_w).
\end{eqnarray}

\hfill\QED

Equipped with the above lemmas, we are ready to demonstrate a contradiction as follows. 
\begin{eqnarray}
L_w &=&L_w/2+L_w/2 \\
&\overset{(\ref{eq:path})(\ref{eq:path2})}{\geq}& H(V_{i_1},\cdots,V_{i_p}|W_{3-k}) + H(V_{i_p},\cdots,V_{i_P}|W_{3-k}) + o(L_w) \\
&\geq&H(V_{i_1},\cdots,V_{i_P}|W_{3-k}) + H(V_{i_p}|W_{3-k}) + o(L_w) \label{eq:lem46}\\
&\overset{(\ref{dec}) (\ref{eq:2color})}{\geq}&H(V_{i_1},V_{i_P}, W_k |W_{3-k}) + L_w/4 + o(L_w) \label{eq:lem47}\\
&\geq&H(W_k|W_{3-k}) + L_w/4 + o(L_w) \\
&\overset{(\ref{h1})}{=}& 5L_w/4  + o(L_w) \\
\Rightarrow ~~ 1 &\geq& 5/4 ~~\mbox{(contradiction)}
\end{eqnarray}
where (\ref{eq:lem46}) follows from submodularity; the first term of (\ref{eq:lem47}) follows from the decoding constraint (\ref{dec}) of the $W_k$-edge $\{V_{i_1}, V_{i_P}\}$ and the second term of (\ref{eq:lem47}) follows by applying Lemma \ref{lemma:2color} to the $2$-color node $V_{i_p}$; the last step follows by dividing by $L_w$ on both hand sides and letting $L_w \rightarrow \infty$.

\subsection{Proof of Theorem \ref{thm:ach}: One Internal Edge} \label{sec:ach}
We show that $R = 4/3$ is achievable if a graph $G(\mathcal{V},\mathcal{E})$ contains only one internal edge, say $W_k$-edge $\{V_i, V_j\}$ and each residing path has at least two special $2$-color nodes. We first present the code construction and then prove it satisfies the decoding constraint (\ref{dec}).

\subsubsection{Code construction}
The first part of the code construction is the same as that in Section \ref{sec:code}. The second part is presented now, where we need to make the following updates. Generate two more generic linear combinations of $W_k$ symbols.
\begin{eqnarray}
\overline{X}_k &=& (\overline{X}_k(1); \overline{X}_k(2)) \triangleq \overline{\bf H}_k W_k  \in \mathbb{F}_p^{2 \times 1} \label{eq:xk}
\end{eqnarray}
where each element of $\overline{\bf H}_k \in \mathbb{F}_p^{2 \times 4}$ is independent and uniform over $\mathbb{F}_p$.

Consider the internal $W_k$-edge $\{V_i, V_j\}$ and find its all residing $W_{3-k}$-paths whose nodes are all $2$-color (i.e., no $1$-color nodes). Suppose there are $M$ such paths, denoted by $P_1, \cdots, P_M$. Start from $V_i \backslash V_j$ and visit each path $P_m, m \in [M]$ along the $W_{3-k}$-edges until we see a special $2$-color node, denoted by $V_{i_m} \backslash V_{j_m}$. Denote the set of $V_{i_m} \backslash V_{j_m}$ nodes as $\mathcal{V}_i \backslash \mathcal{V}_j$. Note that every node in $\mathcal{V}_i, \mathcal{V}_j$ is $W_k$-special and $\mathcal{V}_{i} \cap \mathcal{V}_j = \emptyset$ (as each residing path has at least two special $2$-color nodes).

$V_i, V_j$ are normal and suppose they are currently set as (by the construction in Section \ref{sec:code})
\begin{eqnarray}
     V_i &=& \big( X_k(m^*) + (2j_1-1) X_k(m^* + 1), X_{3-k}(m_1) + (2i_1-1) X_{3-k}(m_1 + 1),  \notag \\
     && ~X_k(m^*) + 2j_1 X_k(m^* + 1) + X_{3-k}(m_1) + 2i_1 X_{3-k}(m_1 + 1) \notag \\
     V_j &=& \big( X_k(m^*) + (2j_2-1) X_k(m^* + 1), X_{3-k}(m_1) + (2i_2-1) X_{3-k}(m_1 + 1),  \notag \\
     && ~X_k(m^*) + 2j_2 X_k(m^* + 1) + X_{3-k}(m_1) + 2i_2 X_{3-k}(m_1 + 1) \big)
\end{eqnarray}
where because the $W_k$-edge $\{V_i, V_j\}$ is internal, the desired symbols are limited to $X_k(m^*), X_k(m^*+1)$. Then each $W_k$-special $2$-color node in $\mathcal{V}_i, \mathcal{V}_j$ is currently set as
\begin{eqnarray}
V \in \mathcal{V}_i \cup \mathcal{V}_j:      V &=& \big( X_k(m^*) + (2j_3-1) X_k(m^* + 1), X_{3-k}(m_2) + (2i_3-1) X_{3-k}(m_2 + 1),  \notag \\
     && ~X_{3-k}(m_2) + 2i_3 X_{3-k}(m_2 + 1) \big) 
\end{eqnarray}
and update it to
\begin{eqnarray}
\mbox{if}~V \in \mathcal{V}_i:      V &=& \big( {\color{red}X_k(m^*)}, X_{3-k}(m_2) + (2i_3-1) X_{3-k}(m_2 + 1),  \notag \\
     && ~X_{3-k}(m_2) + 2i_3 X_{3-k}(m_2 + 1) \big) \label{eq:vi1} \\
\mbox{if}~V \in \mathcal{V}_j:      V &=& \big( {\color{red}X_k(m^*+1)}, X_{3-k}(m_2) + (2i_3-1) X_{3-k}(m_2 + 1),  \notag \\
     && ~X_{3-k}(m_2) + 2i_3 X_{3-k}(m_2 + 1) \big). \label{eq:vj1}
\end{eqnarray}

For every normal $2$-color node $V$ in the segment of residing path $P_m, m \in [M]$ from $V_i$ to the node before $V_{i_m}$. Update $V$ as follows.
\begin{eqnarray}
     V &=& \big( X_k(m^*) + (2j_4-1) X_k(m^* + 1), X_{3-k}(m_3) + (2i_4-1) X_{3-k}(m_3 + 1),  \notag \\
     && ~X_k(m^*) + 2j_4 X_k(m^* + 1) + X_{3-k}(m_3) + 2i_4 X_{3-k}(m_3 + 1) \big)  \\
     \rightarrow~~  V &=& \big( X_k(m^*) + (2j_4-1) {\color{red} \overline{X}_k(1)}, X_{3-k}(m_3) + (2i_4-1) X_{3-k}(m_3 + 1),  \notag \\
     && ~X_k(m^*) + 2j_4 {\color{red} \overline{X}_k(1)} + X_{3-k}(m_3) + 2i_4 X_{3-k}(m_3 + 1) \big). \label{eq:vi2}
\end{eqnarray}
Similarly replace $X_k(m^* + 1)$ by $\overline{X}_k(1)$ for all nodes (except $V_{i_m}$) 
that are connected to the above $V$ through a $W_{3-k}$-path.

For every normal $2$-color node $V$ in the segment of residing path $P_m, m \in [M]$ from $V_j$ to the node before $V_{j_m}$. Update $V$ as follows.
\begin{eqnarray}
     V &=& \big( X_k(m^*) + (2j_5-1) X_k(m^* + 1), X_{3-k}(m_4) + (2i_5-1) X_{3-k}(m_4 + 1),  \notag \\
     && ~X_k(m^*) + 2j_5 X_k(m^* + 1) + X_{3-k}(m_4) + 2i_5 X_{3-k}(m_4 + 1) \big)  \\
     \rightarrow~~  V &=& \big( {\color{red} \overline{X}_k(2)} + (2j_5-1) X_k(m^* + 1), X_{3-k}(m_4) + (2i_5-1) X_{3-k}(m_4 + 1),  \notag \\
     && ~{\color{red} \overline{X}_k(2)} + 2j_5 X_k(m^* + 1) + X_{3-k}(m_4) + 2i_5 X_{3-k}(m_4 + 1) \big). \label{eq:vj2}
\end{eqnarray}
Similarly replace $X_k(m^*)$ by $\overline{X}_k(2)$ for all nodes (except $V_{j_m}$) 
that are connected to the above $V$ through a $W_{3-k}$-path.
The description of the code construction is complete.

\subsubsection{Proof of correctness}
The proof of correctness is similar to that in Section \ref{sec:corr}, where we wish to show that for each edge $\{V_i, V_j\}$, the interference dimension is limited to two so that interference can be decoded and removed and the linear mapping from the four linear combinations of desired source symbols to the four desired source symbols, described by a $4 \times 4$ matrix ${\bf T}_{ij}$ may have full rank and then the existence of a feasible code construction (i.e., a choice of ${\bf H}_1, {\bf H}_2, \overline{\bf H}_k$) is guaranteed by the Schwartz–Zippel lemma \cite{Demillo_Lipton, Schwartz, Zippel} (refer to (\ref{eq:sz})).

We now consider each edge of $G$. The unchanged edges are the same as before and the proof in Section \ref{sec:corr} applies. We are left with the edges that have been updated. 
First, for the only internal $W_k$-edge $\{V_i, V_j\}$, the interference is unchanged, i.e., limited to $X_{3-k}(m_1), X_{3-k}(m_1+1)$ and the desired symbols are $X_k(m^*), X_k(m^*+1), \overline{X}_k(1), \overline{X}_k(2)$ so that $\det({\bf T}_{ij})$ is not the zero polynomial. 
Second, for every $W_{3-k}$-edge in the segment of residing path $P_m, m \in [M]$ from $V_i$ to $V_{i_m}$, the desired $X_{3-k}$ symbols are unchanged and the interference from $W_k$ is limited to $X_k(m^*), \overline{X}_k(1)$, i.e., two dimensions (refer to (\ref{eq:vi1}), (\ref{eq:vi2})) so that interference can be decoded and removed.
Third, for every $W_{3-k}$-edge in the segment of residing path $P_m, m \in [M]$ from $V_j$ to $V_{j_m}$, the desired $X_{3-k}$ symbols are unchanged and the interference from $W_k$ is limited to $X_k(m^*+1), \overline{X}_k(2)$, i.e., two dimensions (refer to (\ref{eq:vj1}), (\ref{eq:vj2})).
Finally, for all other edges that involve a node that has been updated, no matter it is a $W_3$-edge or a $W_{3-k}$-edge, we may verify that interference has dimension two and desired symbols have dimension four. The proof of correctness is thus complete.

\subsection{Proof of Theorem \ref{thm:not_ach}: Graph $G$ in Fig.~\ref{fig:not_ach}} \label{sec:not_ach}
We show that $R < 4/3$ for the graph $G$ in Fig.~\ref{fig:not_ach}.
To set up the proof by contradiction, let us assume that $R = \lim_{L_w\rightarrow \infty}L_w/L_v = 4/3$ is (asymptotically) achievable, i.e., $L_v = 3L_w/4 + o(L_w)$. 

Let us start with a useful inequality, stated in the following lemma.
\begin{lemma}\label{lemma:56}
When $R=4/3$, for the graph $G$ in Fig.~\ref{fig:not_ach}, we have
\begin{eqnarray}
H(V_5,V_6|W_1) \geq L_w/2 + o(L_w). \label{eq:56}
\end{eqnarray}
\end{lemma}

{\it Proof:} 
\begin{eqnarray}
H(V_5,V_6|W_1) 
&=& H(V_1,V_2,V_5,V_6|W_1) - H(V_1,V_2|V_5,V_6,W_1)\\
&\overset{(\ref{dec})}{\geq}& H(V_1,V_2, W_2|W_1) - H(V_1|V_5,W_1) - H(V_2|V_6,W_1) \label{eq:561} \\
&\overset{(\ref{h1})}{\geq}& L_w - H(V_1,V_5|W_1)+H(V_5|W_1) - H(V_2,V_6|W_1)+H(V_6|W_1)\\
&\overset{(\ref{eq:2color}) (\ref{eq:edge})}{\geq}& L_w - L_w/2+ L_w/4 - L_w/2+ L_w/4 + o(L_w) \label{eq:562}\\
& =& L_w/2 + o(L_w)
\end{eqnarray}
where the first term of (\ref{eq:561}) follows from the decoding constraint (\ref{dec}) of the $W_2$-edge $\{V_1, V_2\}$; (\ref{eq:562}) follows by applying Lemma \ref{lemma:edge} to edges $\{V_1, V_5\}, \{V_2, V_6\}$ and applying Lemma \ref{lemma:2color} to nodes $V_5, V_6$.

\hfill\QED

Next, applying Lemma \ref{lemma:edge} to edges $\{V_3, V_5\}, \{V_3, V_6\}, \{V_8, V_5\}, \{V_8, V_6\}, \{V_8, V_7\}, \{V_7, V_4\}$ and submodularity repeatedly, we have
\begin{eqnarray}
3L_w + o(L_w) &\overset{(\ref{eq:edge})}{=}& H(V_3, V_5 | W_1) + H(V_3, V_6 | W_1) + H(V_8, V_5 | W_1) + H(V_8, V_6 | W_1) \notag\\
&&+~H(V_8, V_7 | W_1) + H(V_7, V_4 | W_1) \\
&\geq& H(V_3, V_5, V_6 | W_1) + H(V_3|W_1) + H(V_8, V_5, V_6 | W_1) + H(V_8 | W_1) \notag\\
&&+~H(V_8, V_7, V_4 | W_1) + H(W_7|W_1) \\
&\overset{(\ref{eq:2color})(\ref{eq:normal})}{\geq}& H(V_3, V_4, V_5, V_6, V_7, V_8 | W_1) + H(V_5, V_6 | W_1) + H(V_8|W_1) + 5L_w/4 \notag\\
&& \label{eq:not_ach1} \\
&\overset{(\ref{dec})(\ref{eq:normal})(\ref{eq:56})}{\geq}& H(V_3, V_4, W_2 | W_1) + L_w/2 + L_w/2 + 5L_w/4 \label{eq:not_ach2}\\
&\overset{(\ref{h1})}{\geq}& 13L_w/4 \\
\Rightarrow 3 &\geq& 13/4~\mbox{(contradiction)}
\end{eqnarray}
where (\ref{eq:not_ach1}) follows by applying Lemma \ref{lemma:2color} to $2$-color node $V_7$ and applying Lemma \ref{lemma:normal} to normal $2$-color nodes $V_3, V_8$ and the first term of (\ref{eq:not_ach2}) follows from the decoding constraint (\ref{dec}) of the $W_2$-edge $\{V_3, V_4\}$. We have arrived at a contradiction and the proof is complete.

\subsection{Proof of Theorem \ref{thm:necessary2}: Necessary Condition of $\mathcal{G}_{C=4/3}$ with $K>2$} \label{sec:necessary2}
We show that $R < 4/3$ if a graph $G$ contains any one of the three structures in Theorem \ref{thm:necessary2}. Let us consider the three structures sequentially.

The first structure is that $G$ contains an $M$-color node, where $M \geq 4$. From (\ref{eq:2/3}), we have $R \leq (M+1)/M \leq 5/4 < 4/3$. 

The second structure is that $G$ contains a $3$-color node $V$ that is connected to an $M$-color node $V_{i_1}$, where $M \geq 2$. Suppose $\{V, V_{i_1}\}$ is a $W_{k_1}$-edge. As $V$ is $3$-color, we have a $W_{k_2}$-edge $\{V, V_{i_2}\}$ and a $W_{k_3}$-edge $\{V, V_{i_3}\}$, where $k_1, k_2, k_3$ are distinct and $i_1, i_2, i_3$ are distinct. As $V_{{i}_1}$ is $M$-color, $M \geq 2$, we have a $W_k$-edge $\{V_{i_1}, V_j\}$ where $j$ might be $i_2$ or $i_3$ (but $j \neq i_1$) and $k$ might be $k_2$ or $k_3$ (but $k \neq k_1$). The following proof will work under all circumstances.

Consider $W_{k_2}$-edge $\{V, V_{i_2}\}$ and $W_{k_3}$-edge $\{V, V_{i_3}\}$. From the decoding constraint (\ref{dec}), we have
\begin{eqnarray}
2L_w &\overset{(\ref{h1})(\ref{dec})}{=}& I(V, V_{i_2}, V_{i_3}; W_{k_2}, W_{k_3}) \\
&=& H(V, V_{i_2}, V_{i_3}) - H(V, V_{i_2}, V_{i_3} | W_{k_2}, W_{k_3}) \\
&\leq& 3 L_v - H(V | W_{k_2}, W_{k_3}) \\
\Rightarrow H(V | W_{k_2}, W_{k_3}) &\leq& 3 L_v - 2 L_w. \label{eq:necessary21}
\end{eqnarray}

Consider $W_k$-edge $\{V_{i_1}, V_j\}$. From (\ref{eq:v1lw}), we have
\begin{eqnarray}
H(V_{i_1} | W_k) \leq 2L_v - L_w. \label{eq:necessary22}
\end{eqnarray}

Adding (\ref{eq:necessary21}) and (\ref{eq:necessary22}), we have
\begin{eqnarray}
5L_v - 3L_w &\geq& H(V | W_{k_2}, W_{k_3}) + H(V_{i_1} | W_k) \\
&\geq&  H(V, V_{i_1} | W_{k_2}, W_{k_3}, W_k) \\
&\overset{(\ref{dec})}{\geq}& H(W_{k_1} | W_{k_2}, W_{k_3}, W_k) \\
&\overset{(\ref{h1})}{=}& L_w \\
\Rightarrow~~ R= L_w/L_v &\leq& 5/4 < 4/3.
\end{eqnarray}

The third structure is that $G$ contains a normal $2$-color node $V$ that is connected to a $2$-color node $V_i$ and $V, V_i$ are connected to different types of edges. Suppose $\{V, V_i\}$ is a $W_k$-edge. As $V, V_i$ are $2$-color (the two colors are different) and $V$ is normal, we have a $W_{k_1}$-edge $\{V_i, V_{i_1}\}$, 
a $W_{k_2}$-edge $\{V, V_{j_1}\}$, and a $W_{k_3}$-edge $\{V_{j_1}, V_{j_2}\}$ where $k, k_1, k_2$ are distinct and $k_3\neq k_2$. 

Consider $W_k$-edge $\{V, V_{i}\}$ and $W_{k_1}$-edge $\{V_{i},V_{i_1}\}$. Following the derivation of (\ref{eq:necessary21}), we have
\begin{eqnarray}
H(V |W_k,W_{k_1}) \leq 3L_v-2L_w. \label{eq:necessary23}
\end{eqnarray}

Consider $W_{k_3}$-edge $\{V_{j_1}, V_{j_2}\}$. From (\ref{eq:v1lw}), we have
\begin{eqnarray}
H(V_{j_1}|W_{k_3}) \leq 2L_v-L_w. \label{eq:necessary24}
\end{eqnarray}

Adding (\ref{eq:necessary23}) and (\ref{eq:necessary24}), we have
\begin{eqnarray}
5L_v - 3L_w &\geq& H(V | W_{k}, W_{k_1}) + H(V_{j_1} | W_{k_3}) \\
&\geq&  H(V, V_{j_1} | W_{k}, W_{k_1}, W_{k_3}) \\
&\overset{(\ref{dec})}{\geq}& H(W_{k_2} | W_{k}, W_{k_1}, W_{k_3}) \\
&\overset{(\ref{h1})}{=}& L_w \\
\Rightarrow~~ R= L_w/L_v &\leq& 5/4 < 4/3.
\end{eqnarray}

\subsection{Proof of Theorem \ref{thm:sufficient2}: Sufficient Condition of $\mathcal{G}_{C=4/3}$ with $K>2$}\label{sec:sufficient2}
We show that $R=4/3$ is achievable if a graph $G(\mathcal{V},\mathcal{E}) \notin \mathcal{G}_{C<3/4}^{\mbox{\tiny Thm \ref{thm:necessary2}}}$ contains no internal edge after removing one $1$-color nodes in residing paths. We first present the code construction, which is a minor modification of that in Section \ref{sec:code} and then prove it satisfies the decoding constraint (\ref{dec}), which is similar to that in Section \ref{sec:corr}.

\subsubsection{Code construction}
Choose the field size $p$ to be a prime that is greater than $4|\mathcal{E}|$. 
Set $L_w = 4\log_2 p, L_v = 3 \log_2 p$ so that each source$\backslash$coded symbol is comprised of $4\backslash 3$ symbols from $\mathbb{F}_p$ and the rate achieved is $4/3$. 

Consider the set of nodes that are connected to $W_k$-edges, $k \in [K]$ and denote this set by $\mathcal{V}_k$. 
Consider the subgraph of $G(\mathcal{V}, \mathcal{E})$ whose node set is $\mathcal{V}_k$ 
and edge set is comprised of all edges that are not $W_k$-edges and are connected to some node in $\mathcal{V}_k$, denoted by $\mathcal{E}_{k^c}$ and denote this subgraph by ${G}_k(\mathcal{V}_k, \mathcal{E}_{k^c})$. Decompose ${G}_k(\mathcal{V}_k, \mathcal{E}_{k^c})$ into $W_{k'}$-components, $k' \neq k$ and suppose we have $M_k$ such components. A trivial component with a single node can be classified as a $W_{k'}$-component for any $k' \neq k$ and we just fix one $k'$ (any choice will work).
Among these $M_k$ $W_{k'}$-components, suppose $M_k^1$ components are comprised of $1$-color nodes and label them as $P_k^{[1]}, \cdots, P_k^{[M_k^1]}$; $M_k^2$ components are comprised of $2$-color nodes and label them as $Q_k^{[1]}, \cdots, Q_k^{[M_k^2]}$; the remaining $M_k^3 = M_k - M_k^1 - M_k^2$ components are comprised of $3$-color nodes and label them as $S_k^{[1]}, \cdots, S_k^{[M_k^3]}$ (each such component is an isolated node as $3$-color nodes are connected to only $1$-color nodes when $G\notin \mathcal{G}_{C<3/4}^{\mbox{\tiny Thm \ref{thm:necessary2}}}$). 

Generate generic linear combinations of the source symbols as follows.
\begin{eqnarray}
W_k &=& (W_k(1); \cdots; W_k(4)) \in \mathbb{F}_{p}^{4 \times 1}, k \in [K] \\
X_k &=& (X_k(1); \cdots; X_k(3M_k^1 + 2M_k^2 + M_k^3)) \triangleq {\bf H}_k W_k  \in \mathbb{F}_p^{(3M_k^1 + 2M_k^2 + M_k^3) \times 1} \label{eq:xkz}
\end{eqnarray}
where each element of ${\bf H}_k \in  \mathbb{F}_p^{(3M_k^1 + 2M_k^2 + M_k^3) \times 4}$ is chosen uniformly and independently from $\mathbb{F}_p$. 

Consider $P_k^{[m]}, m \in [M_k^1]$, denote its node by $V$, and set
\begin{eqnarray}
V = \big(X_k(3m-2), X_k(3m-1), X_k(3m) \big). \label{eq:1z}
\end{eqnarray}
This step completes the assignment for all $1$-color nodes.

Consider $Q_k^{[m]}, m \in [M_k^2]$. Suppose $Q_k^{[m]}$ contains $J$ nodes $V_{i_1}, \cdots, V_{i_J}$. Consider $V_{i_j}, j \in [J]$, which is $2$-color.
\begin{eqnarray}
&& \mbox{If $V_{i_j}$ is $W_k$-special, set $V_{i_j}^{[k]} \triangleq X_k\big(3M_k^1 + 2m-1 \big) + (2j-1) X_k\big(3M_k^1 + 2m \big) $;} \\
&& \mbox{otherwise, set $V_{i_j}^{[k]} = (V_{i_j}^{[k]}(1), V_{i_j}^{[k]}(2)) \triangleq \Big( X_k\big(3M_k^1 + 2m-1 \big) + (2j-1) X_k\big(3M_k^1 + 2m \big),$} \notag\\
&& ~~~~~~~~~~~~~~~~~~~~~~~~~~~~~~~~~~~~~~~~~~~~~~~~~~~~~~~\mbox{$X_k\big(3M_k^1 + 2m-1 \big) + 2j X_k\big(3M_k^1 + 2m \big) \Big) $.} \label{eq:2sz}
\end{eqnarray}
For any $2$-color node $V$ that is connected to $W_{k_1}$-edges and $W_{k_2}$-edges,
\begin{eqnarray}
&& \mbox{if $V$ is normal, then set $V = (V^{[k_1]}(1), V^{[k_2]}(1), V^{[k_1]}(2) + V^{[k_2]}(2))$;} \\
&& \mbox{otherwise, set $V = (V^{[k_1]}, V^{[k_2]})$.} \label{eq:2z}
\end{eqnarray}
This step completes the assignment for all $2$-color nodes. 

Consider $S_k^{[m]}, m \in [M_k^3]$, denote its node by $V$, and set
\begin{eqnarray}
V^{[k]} \triangleq X_k(3M_k^1 + 2M_k^2 + m).
\end{eqnarray}
For any $3$-color node $V$ that is connected to $W_{k_1}$-edges, $W_{k_2}$-edges, and $W_{k_3}$-edges, set
\begin{eqnarray}
V = (V^{[k_1]}, V^{[k_2]}, V^{[k_3]}). \label{eq:3z}
\end{eqnarray}
This step completes the assignment for all $3$-color nodes and as $G \notin \mathcal{G}_{C<3/4}^{\mbox{\tiny Thm \ref{thm:necessary2}}}$ contains no $M$-color nodes, where $M \geq 4$, the code construction is complete.

\subsubsection{Proof of correctness}
Consider any edge $\{V_i, V_j\} \in \mathcal{E}$ and suppose it is a $W_k$-edge.

When $V_i, V_j$ contain one $1$-color code, then our assignment ensures that $(V_i, V_j)$ contains four distinct elements of $X_k$, which can be written as ${\bf T}_{ij}^{4\times 4} W_k$ and $\det({\bf T}_{ij})$ is a non-zero polynomial $T_{ij}({\bf H}_1, \cdots, {\bf H}_K)$.

We are left with cases where $V_i, V_j$ are both $2$-color (because $3$-color nodes in $G \notin \mathcal{G}_{C<3/4}^{\mbox{\tiny Thm \ref{thm:necessary2}}}$ are connected only to $1$-color nodes). 
When $V_i, V_j$ contain one normal $2$-color code, then $G \notin \mathcal{G}_{C<3/4}^{\mbox{\tiny Thm \ref{thm:necessary2}}}$ ensures that $V_i, V_j$ are connected to edges that are associated with the same set of two source symbols, i.e., we are back to the $K=2$ setting considered in Theorem \ref{thm:sufficient} and following the proof in Case 2 and Case 3 of Section \ref{sec:corr}, we have $T_{ij}({\bf H}_1, \cdots, {\bf H}_K)$ is non-zero. The only remaining case is that $V_i, V_j$ are both special, say $V_i$ is $W_{k_1}$-special and $V_j$ is $W_{k_2}$-special, where $k_1 \neq k, k_2 \neq k$. Then from (\ref{eq:2sz}) and (\ref{eq:2z}), we know that $(V_i, V_j)$ each contains two distinct elements of $X_k$ (distinctness is due to the absence of internal edges after removing $1$-color nodes) so that $T_{ij}({\bf H}_1, \cdots, {\bf H}_K)$ is not the zero-polynomial.

Finally, consider $\prod_{i,j: \{V_i, V_j\} \in \mathcal{E}} T_{ij}({\bf H}_1, \cdots, {\bf H}_K)$, which is a non-zero polynomial with degree at most $4 |\mathcal{E}| < p$, the field size. By the Schwartz–Zippel lemma \cite{Demillo_Lipton, Schwartz, Zippel}, there exists a realization of ${\bf H}_1, \cdots, {\bf H}_K$ so that each $T_{ij}({\bf H}_1, \cdots, {\bf H}_K) \neq 0$ and all decoding constraints (\ref{dec}) are satisfied.

\section{Discussion}
An extremal rate perspective is taken to study the storage code problem over graphs. For the highest capacity values, we have identified a number of combinatorial structures that have significant impact on the code rate - $M$-color code (i.e., the number of sources associated with a node), internal edge (which captures a direct conflict between alignment of undesired source symbols and independence of desired source symbols), normal $2$-color node$\backslash$special $2$-color node (for rate 4/3, which keeps the same interference$\backslash$which could change interference up to the extent of $1/4$ source size). Both the achievability and converse results are guided by a linear dimension counting view. The sufficient and necessary conditions presented are not the largest that our proof technique can lead to, i.e., we can solve more graph instances, but a systematic description is still out of current reach. It is not clear which rates will turn out to have hard capacity instances. Specifically, all extremal graphs with storage code capacity $4/3$ appear to go beyond the techniques of this work. Regarding generalizations, we note that our model is the most elementary, where we have focused on the highest capacity values, i.e., best rate scenarios instead of lowest capacity values, i.e., worst rate scenarios, or other physically meaningful rates; decoding constraints are placed on a pair of nodes in this work instead of an arbitrary set of nodes, i.e., we may have a hypergraph rather than a graph \cite{Sahraei_Gastpar}; each edge is associated with only one source symbols instead of multiple source symbols where the decoding structure can be more diverse \cite{Li_Sun_SecureStorage}.
Finally, from an extremal rate and network perspective, we may view combinatorial objects using the metric of capacity and study further extremal (largest, densest, most (linearly) independent) graphs, set families, vector spaces etc. along the line of extremal combinatorics \cite{Jukna_Extremal}.

\let\url\nolinkurl
\bibliographystyle{IEEEtran}
\bibliography{Thesis}
\end{document}